\pdfoutput=1
\documentclass[aps,pre,twocolumn,groupedaddress,superscriptaddress,showpacs,longbibliography]{revtex4-1}
\usepackage{amsmath}
\usepackage{hyperref} 
\usepackage{cases}
\usepackage{graphicx}
\usepackage{bbm}
\usepackage{color}
\usepackage{soul}
\usepackage{dsfont}
\usepackage{enumitem}
\usepackage{relsize}
\usepackage{mathtools}
\usepackage{latexsym}

\newcommand{\m}{\ensuremath{\mu}}
\renewcommand{\a}{\ensuremath{\alpha}}
\renewcommand{\b}{\ensuremath{\beta}}
\renewcommand{\c}{\ensuremath{\gamma}}
\newcommand{\g}{\ensuremath{\gamma}}

\renewcommand{\d}{\ensuremath{\delta}}
\renewcommand{\t}{\ensuremath{\tau}}

\newcommand{\til}[1]{\tilde{#1}}

\newcommand{\off}{\ensuremath{\mbox{\small OFF}}}
\newcommand{\on}{\ensuremath{\mbox{\small ON}}}
\newcommand{\OFF}{{\small OFF }}
\newcommand{\ON}{{\small ON }}
\newcommand{\Prob}{\ensuremath{\mbox{\small Prob}}}

\begin{document}

\title{Temporally correlated zero-range process with open boundaries:\\Steady state and fluctuations}

\author{Massimo Cavallaro}
\email[]{m.cavallaro@qmul.ac.uk}
\affiliation{School of Mathematical Sciences, Queen Mary University of London, Mile End Road, London, E1 4NS, UK}
\author{Ra\'ul J. Mondrag\'on}
\email[]{r.j.mondragon@qmul.ac.uk}
\affiliation{School of Electronic Engineering and Computer Sciences, Queen Mary University of London, Mile End Road, London, E1 4NS, UK}
\author{Rosemary J. Harris}
\email[]{rosemary.harris@qmul.ac.uk}
\affiliation{School of Mathematical Sciences, Queen Mary University of London, Mile End Road, London, E1 4NS, UK}
\begin{abstract}
We study an open-boundary version of the on-off zero-range process introduced
in Hirschberg \textit{et al.} [Phys.\ Rev.\ Lett.\ \textbf{103}, 090602 (2009)].
This model includes temporal correlations which can promote the condensation of particles,
a situation observed in real-world dynamics.
We derive the exact solution for the steady state of the one-site system,
as well as a mean-field approximation for larger one-dimensional lattices,
and also explore the large deviation properties of the particle current.
Analytical and numerical calculations show that, although the particle distribution is well described by an effective Markovian solution,
the probability of rare currents differs from the memoryless case.
In particular, we find evidence for a memory-induced dynamical phase transition.
\end{abstract}


\maketitle
\section{Introduction}
The first step in the study of a complex system is often to focus on its typical behaviour.
Employing the tools of statistical mechanics, for example,
we can study the typical properties of a macroscopic system.
However, there are situations in which the behaviour of interest is not typical, but rather atypical.
For example, the transport of energy, particles or vehicles could be enhanced by exceptional
coherent configurations, or occasionally delayed when an instantaneous situation similar to condensation occurs~\cite{schadschneider2010stochastic}.
As another example, in communication networks it is very important to predict how likely it is to have interruptions or packet loss~\cite{Smith2011}.

Moreover, rare events help to shed light on the foundations of non-equilibrium statistical mechanics,
just as they play an important role in defining the thermodynamic potentials in equilibrium
statistical mechanics~\cite{Ellis1995,Ellis2005,Touchette2009}.
While a general framework for the characterisation of systems far from thermal equilibrium is at a primitive stage,
large deviation theory plays a central role~\cite{Touchette2009,Touchette2012}.

As a comprehensive theory of non-equilibrium phenomena does not exist,
the analytical study of toy models is an effective way to build it up.
The majority of the interacting-particle models in the literature are Markovian, i.e. memoryless.
Such an approximation simplifies the theoretical treatment, but can exclude some properties of physical phenomena.
The effects of memory on such models have thus prompted recent curiosity~\cite{Hirschberg2009,Hirschberg2012,Concannon2014,Khoromskaia2014}.
We enter into this context by studying a driven-diffusive  system
which is referred to as the \textit{on-off} zero-range process (on-off ZRP) and focusing on its non-equilibrium aspects.
This model is an open-boundary version of the non-Markovian ZRP defined in~\textcite{Hirschberg2009,Hirschberg2012}
and allows analytical progress.

Non-equilibrium stationary states (NESSs) are characterised by the presence of finite currents,
which measure the violation of detailed balance for opposing transitions
between two configurations~\cite{Zia2007,Qian2010,Platini2011}.
Such currents fluctuate in time and the functions
that determine the probability of deviation from their typical values
have the same mathematics as the thermodynamical functionals
defined in equilibrium statistical mechanics.
In this spirit, we are interested in the particle current for our model,
i.e., in the transition events corresponding to particle hops.
The study of its rare fluctuations reveals effects
of the time-correlation hidden in the stationary state.

The paper is organised as follows.
In Sec.~\ref{sec:model}, we define the model and in
Sec.~\ref{sec:stationary_distribution} we derive its 
stationary state in the single-site system.
In Sec.~\ref{sec:mean_field}, we present a mean-field treatment of the
dynamics on a chain topology and test the validity of this approximation against standard Monte Carlo simulations.
In Sec.~\ref{sec:fluctuations}, we explore the current fluctuations,
focusing, in a one-site system, on the difference between the small fluctuation regime, obtained by
analytic continuation of the NESS~(Sec.~\ref{sec:small_fluctuations}),  and the extreme fluctuation
regimes (Sec.~\ref{sec:large_current_fluctuations}), and deriving the phase boundaries between them (Sec.~\ref{sec:range_of_validity}).
The analytical results are tested  against an advanced numerical method which has been developed to evaluate
large deviation functions directly~\cite{Lecomte2007}.
We summarize the results in Sec.~\ref{sec:conclusions}.

\section{Model}
\label{sec:model}
The ZRP is a model of interacting particles on a discrete lattice,
which we take here to be a one-dimensional chain.
Each lattice site can contain an arbitrary positive number of particles.
The evolution proceeds  in continuous time, i.e., transitions occur after a waiting time
which is an exponentially distributed random variable.
Specifically, in the standard ZRP, a particle can hop to one of the neighbouring sites with rate proportional to $\m_n$,
which depends only on the  occupation number $n$ of the departure site.
Obviously, the departure rate from an empty site is given by $\m_0 = 0$.
The functional form of $\m_n$ encodes the interaction between particles,
which occurs only on the departure site, hence the epithet zero-range.  The special case $\m_n = a n$, where $a>0$ is a constant,
corresponds to free particles  since in this case each particle leaves the site independently from the others.
Other choices of $\m_n$ correspond to attractive or repulsive inter-particle interaction
if the $n$-dependence is sublinear or superlinear, respectively.

Models with zero-range interactions have proven to display complex collective behaviour
whilst allowing analytical treatment~\cite{Evans2005}.
In particular, the ZRP is well suited for theoretical analysis because the
stationary probability distribution of a given configuration factorises and can be
calculated exactly~\cite{Spitzer1970}.
It is worth mentioning that certain choices of $\m_n$ lead to condensation,
i.e., the accumulation of a macroscopic fraction of the total number of particles on a single site.
Condensation transitions far from equilibrium have been studied in physics~\cite{Eggers1999},
as well as in economics~\cite{Bouchaud2000,Burda2002}, biology~\cite{Frohlich1975},
network science~\cite{Bianconi2001,DeMartino2009}, and queueing theory~\cite{Chernyak2010}.
Toy models, such as the ZRP, provide a theoretical foundation
for understanding condensation in these systems.
Exact results from the ZRP have also been used in models of vehicular traffic~\cite{Kaupuzs2005,schadschneider2010stochastic},
reptation in polymer physics~\cite{Antal2009}, and
transport and coalescence in granular systems~\cite{Torok2005}.

A further step towards a deeper comprehension of real-world phenomena may be achieved by
studying  stochastic systems with time correlations.
A modified zero-range process with non-Markovian dynamics has been introduced in~\citet{Hirschberg2009,Hirschberg2012}.
The crucial new ingredient is that each site has an additional \textit{clock/phase} variable $\t$ and the particles
cannot leave the site when the clock is set to zero, which corresponds to the \OFF phase.
The clock ticks and turns on with rate $c$ and turns off with each particle arrival.
This mechanism favours the accumulation of particles on a site.
According to the zero-range dynamics, the particles interact only on site,
but now have a different departure rate $\mu_{n,\t}$. The additional variable
$\t$ takes into account events in particle configuration space that happened in the past
and therefore introduces temporal correlations.
This model has sparked interest as it displays, under certain conditions,
a condensate with slow drift~\cite{Hirschberg2009,Hirschberg2012}.
Systems with distinct on and off phases are also of interest as models
for intra-cellular ion-channels~\cite{VanDongen2004,Mitra2014} and for data traffic streams~\cite{Mondragon2001},
as well as providing examples of stochastic processes with non-convex rate functions~\cite{Duffy2008}.

\begin{figure}
 \includegraphics[width=8cm]{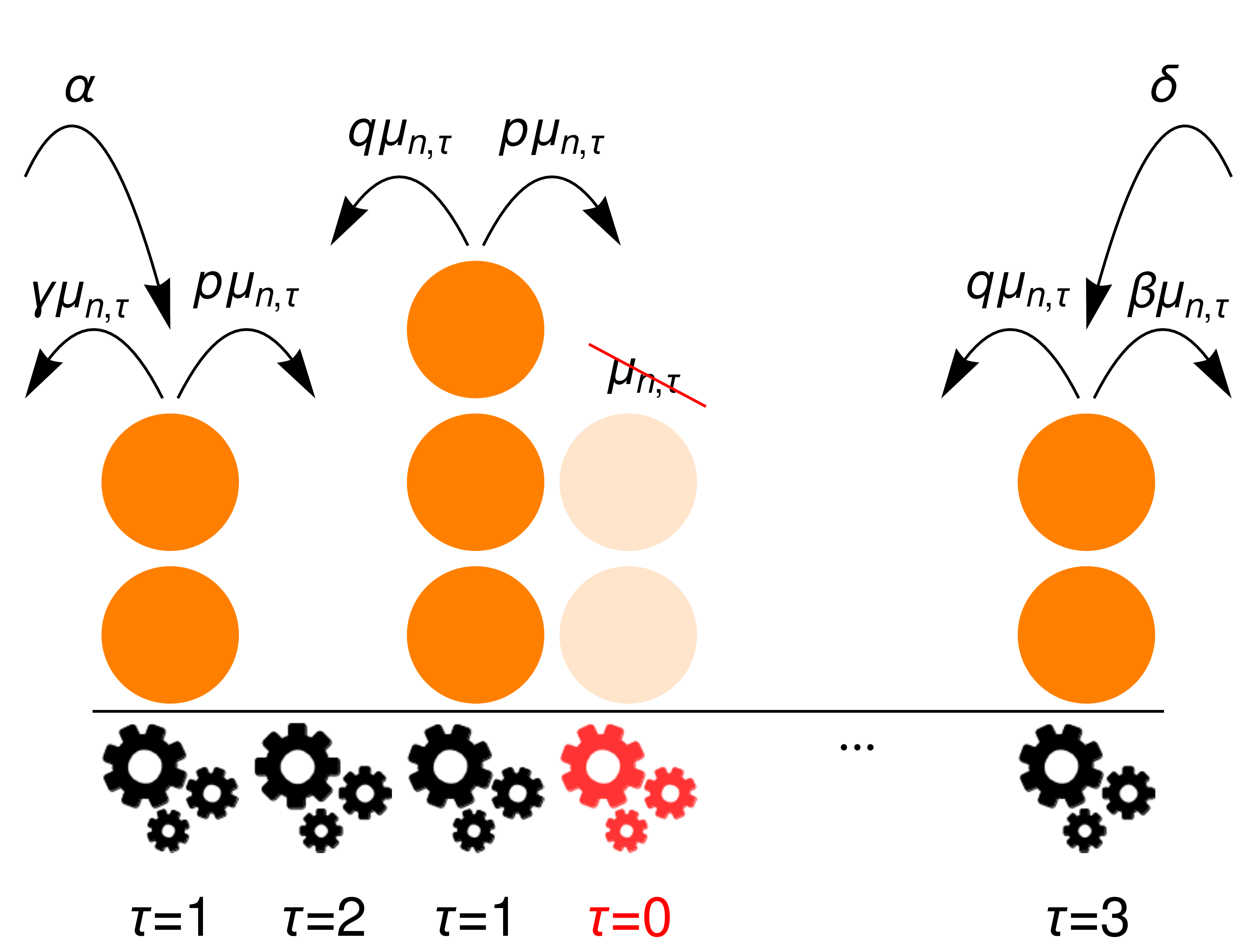}
 \caption{  \label{fig:ZRPimagePA} (Color online) Non-Markovian ZRP on a one-dimensional lattice with open boundaries.
 Each site has a hidden variable $\t$, whose values are represented by the positions of a gear, which controls the departure rate.
 When $\t$ assumes   value zero no departure is possible and the corresponding state is referred to as~\off.
 This lock-down occurs in conjunction with the arrival of a particle. 
 }
\end{figure}

To the best of our knowledge, the ZRP with on-off dynamics has been studied only on ring topology, i.e., with periodic boundary conditions.
In this paper we investigate the open-boundary version of the model, thus extending the work of~\textcite{Hirschberg2009,Hirschberg2012}.
We implement the same dynamics on an open chain with arbitrary hopping rates and boundary parameters,
see Fig.~\ref{fig:ZRPimagePA}.
Particles are added and removed through the boundaries.
On the leftmost lattice site (site $1$), particles are injected with rate $\a$ and they are removed
with rate $\g \m_{n,\tau}$ which is non-zero only when the phase of site $1$ is different from $\t=0$.
Similarly, on the rightmost site (site $L$) particles are removed with rate  $\b \m_{n,\tau}$,
according to the phase of site $L$, and are injected with rate $\d$.
This situation corresponds to a bulk system in contact with two different reservoirs.
In the bulk, particles jump to the left (right) with rate $q \mu_{n,\t}$ ($p \mu_{n,\t}$),
which is again non-zero only when the phase of the departure site is not $\t=0$.
The dynamics is sensitive to the specific rate values and we consider choices that induce a rightwards driving along the chain.
In particular, it is worth making the distinction between the totally asymmetric (TA) and the partially asymmetric (PA) processes.

Hereafter, we will consider explicitly two forms for interaction factor $\m_{n,\t}$,
i.e., the case where $\m_{n,\t}$ is constant with respect to $n>0$, which corresponds to an on-site attractive interaction between particles,
and the case where $\m_{n,\t}$ is linear in $n$, which corresponds to no direct interaction between particles (excluding residual
correlations due to the blockade mechanism).

\begin{figure}
 \centering
 \includegraphics[width=7.5cm]{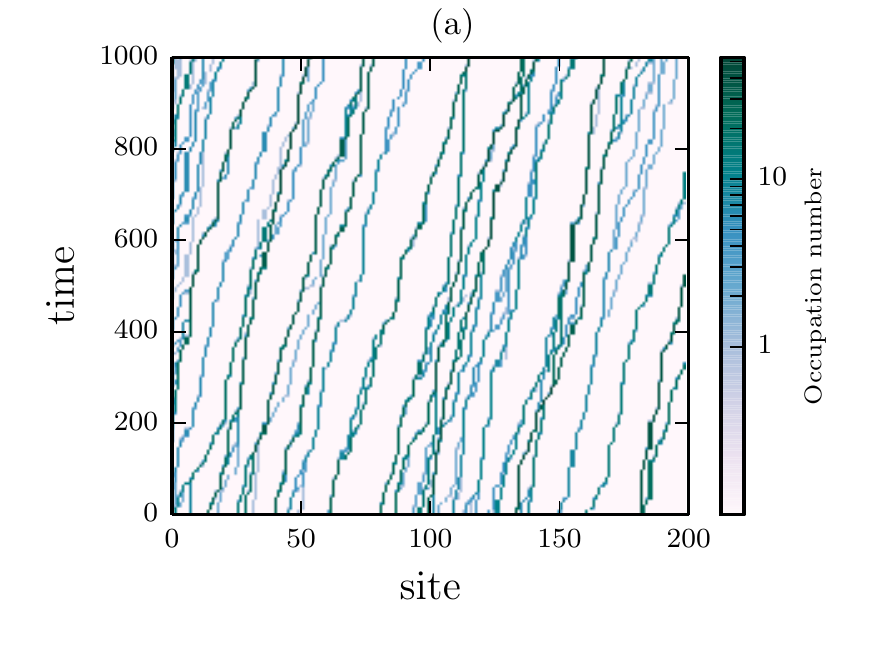}
 \includegraphics[width=7.5cm]{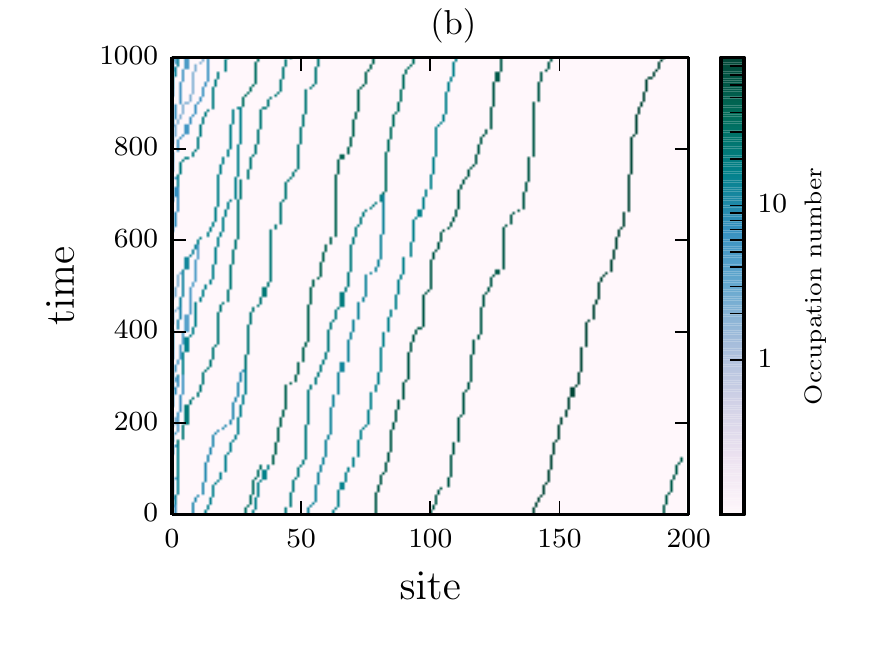}
 \includegraphics[width=7.5cm]{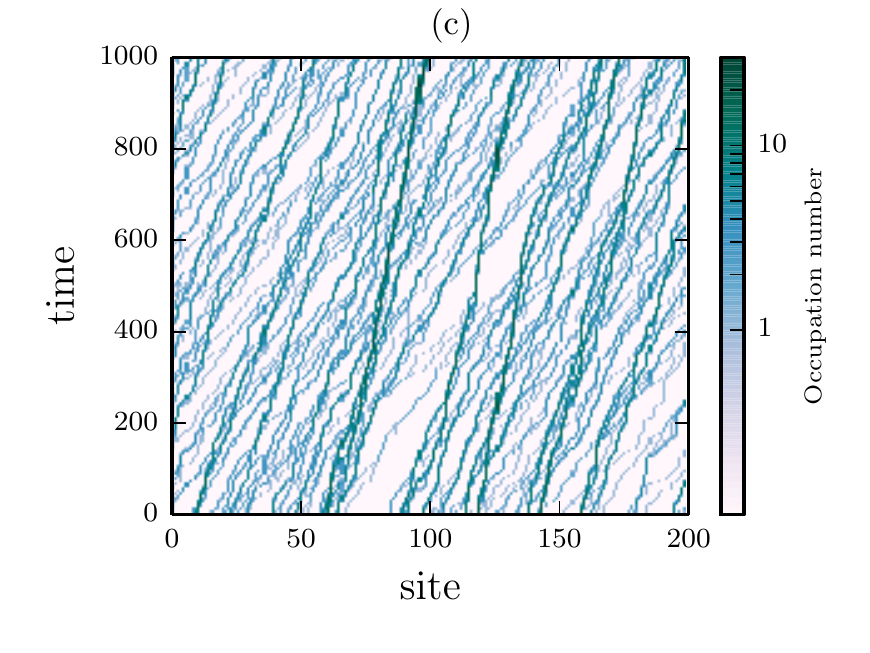}
 \caption{\label{fig:density profile evolution} (Color online) Monte Carlo
          time evolution of the occupation profile of the on-off ZRP on a
          one-dimensional lattice. (a) Rates $\mu_{n,\t}=n$ if $\t>0$, $\mu_{n,\t}=0$ otherwise,
          and  $(\a,\b,\g,\d,p,q,c) = (0.1,0.2,0,0,1,0,0.05)$.
          Only correlations due to the blockade mechanism are present.
          The particles organise in travelling clusters.
          Their speed is mainly  governed by $c$.
          (b) Same parameters as the former case, except $(\b,p) = (10^4,10^4)$.
          The particles jump almost simultaneously to the next site
          as soon as the blockade is removed. Each particle cluster tends to occupy a single site.
          The drift proceeds with a rate $\simeq c$.
          (c) Same parameters as (a), except $\mu_{n,\t}=1$ if $\t>0$ and $n>0$, $\mu_{n,\t}=0$ otherwise, and $c=0.15$. 
          As a result of the attractive inter-particle interaction,
          the clusters with more particles travel slower than the less populated ones.
          This mechanism enhances congestion.
          }
\end{figure}

The stationary distribution of the standard ZRP on an open chain
has been extensively studied in~\textcite{Levine2005}. 
In this case the particles are distributed along the system according to
a product-form structure that implies no correlations between sites.
In contrast, the on-off ZRP can generate more complex patterns,
as shown in Fig.~\ref{fig:density profile evolution} for three sets of parameters.
The clock-tick rate $c$ plays a major role in these patterns, the lower its value is, the more important the correlations are.
Increasing the value of $c$, the system eventually becomes spatially uncorrelated.
In the next two sections we study in detail how the introduction of time correlations affects the stationary state
and the current fluctuations.

\section{Stationary state}
\subsection{Exact results for one-site system}
\label{sec:stationary_distribution}
As mentioned above, a notable property of the standard ZRP is that in the stationary state the probability $ P^*(\{n_l\}) $ of finding the system
in a state $\{n_l\} = (n_1,n_2,\dots n_L)$ with $n_l$ particles on site $l$,  is given by a simple factorised form
\begin{equation}
    P^*(\{n_l\}) = \prod^L_{l=1} {P}^*_l (n_l),
    \label{eq:product_solution}
\end{equation}
where  $P^*_l(n_l)$ is the probability of finding the site $l$ with $n_l$ particles.
The one-site marginals are determined by
\begin{equation}
    P_l^*(n) =  \frac{{z_l}^n}{Z_l} \prod_{i=1}^{n} \m_i^{-1},
    \label{eq:stationary_uno}
\end{equation}
where $z_l$ is a site-dependent fugacity (which is a function of the hopping rates) and
the grand-canonical partition function $Z_l = \sum^\infty_{n=0} {z_l}^n \prod_{i=1}^{n} \m^{-1}_i$
ensures normalisation~\cite{Levine2005}. It is worth noting that, for
certain choices of $\mu_n$, it is not possible for the sum in $Z_l$ to converge for all $z_l$.
The divergence of $Z_l$ corresponds to the accumulation of particles on the site $l$ and we refer to it as congestion.
Indeed, the infinite accumulation on one or more sites in an open system 
can be thought of as a kind of condensation phenomenon~\cite{Levine2005,Chernyak2010}.
In the following, we will also use the ``condensation'' terminology even for the single-site case.

Our preliminary simulations in Fig.~\ref{fig:density profile evolution} suggest that 
we cannot rely on a factorised steady state for the non-Markovian model introduced in Sec.~\ref{sec:model}.
However, for the single site system, an exact solution is straightforward.
The state is defined by two variables:
the number of particles in the box $n$ and a ``clock'' or ``phase'' variable $\tau$.
We focus on TA dynamics and consider a box which receives particles with rate $\a$ and ejects
particles with rate $\b \mu_{n,\t}$, where $\mu_{n,\t}$ is a function of the box state.
The departure event is possible only when $\t \neq 0$.
Also, the dynamics includes the advance of the clock with rate $c$,
and the reset to $\tau=0$ when a particle arrives.
If one defines $P(-1,\tau;t)=P(n,-1;t)=0$,
the following Master equation is valid for $\tau \ge 0$ and $ n \ge 0$:
\begin{multline}
	\frac{d P(n,\tau;t)}{dt} = c P(n,\tau -1;t) + \b \mu_{n+1,\t} P(n+1,\tau;t) \\
	+ \delta_{\tau,0} \sum_{\tau' \ge 0} \a  P(n-1, \tau';t) 
	- (c +  \b \m_{n,\tau} + \a) P(n,\tau;t) ,
\label{eq:hirsch_nt}
\end{multline}
where $P(n,\tau;t)$ denotes the probability of finding the system with $n$ particles
and phase $\tau$ at time $t$ and $\delta_{\tau,0}$ is a Kronecker delta.
The first term on the right-hand side of~\eqref{eq:hirsch_nt}
corresponds to a clock tick, the second term to the departure of a particle,
the third term to the arrival of a particle and the fourth term to the respective
escape events from the state $(n,\tau)$.

As in~\textcite{Hirschberg2009,Hirschberg2012}, we choose to simplify the dependence
of the jump rate on $\tau$ to $\m_{n,\t} = \m_n$ when $\tau>0$. 
Hereafter we specialise to this case, except when we explicitly refer to a general form for $\m_{n,\t}$.
In this simplified case, it is more convenient to write the Master equation~\eqref{eq:hirsch_nt} in terms of
$P(n,\on;t) = \sum_{\t>0} P(n,\t;t)$ and $P(n,\off;t) = P(n,0;t) $
 \begin{align}
  \frac{dP(n,\on;t)}{dt}=& c P(n,\off;t) + \b \m_{n+1}  P(n+1,\on;t)\notag \\
  &- (\b \m_n + \a )P(n,\on;t), \label{eq:1}\\
  \frac{dP(0,\on;t)}{dt}=&  c P(0,\off;t) + \b \mu_1 P(1,\on;t) \notag\\
    &-   \a P(0,\on;t), \label{eq:2} \\
  \frac{dP(n,\off;t)}{dt}=& \a P(n-1,\on,t)+\a P(n-1,\off,t) \notag \\
  &- ( c+\a ) P(n,\off;t), \label{eq:3} \\
  \frac{dP(0,\off;t)}{dt}= &- (c+\a ) P(0,\off;t). \label{eq:4}
 \end{align}

\begin{figure}
    \centering
    \includegraphics[width=8cm]{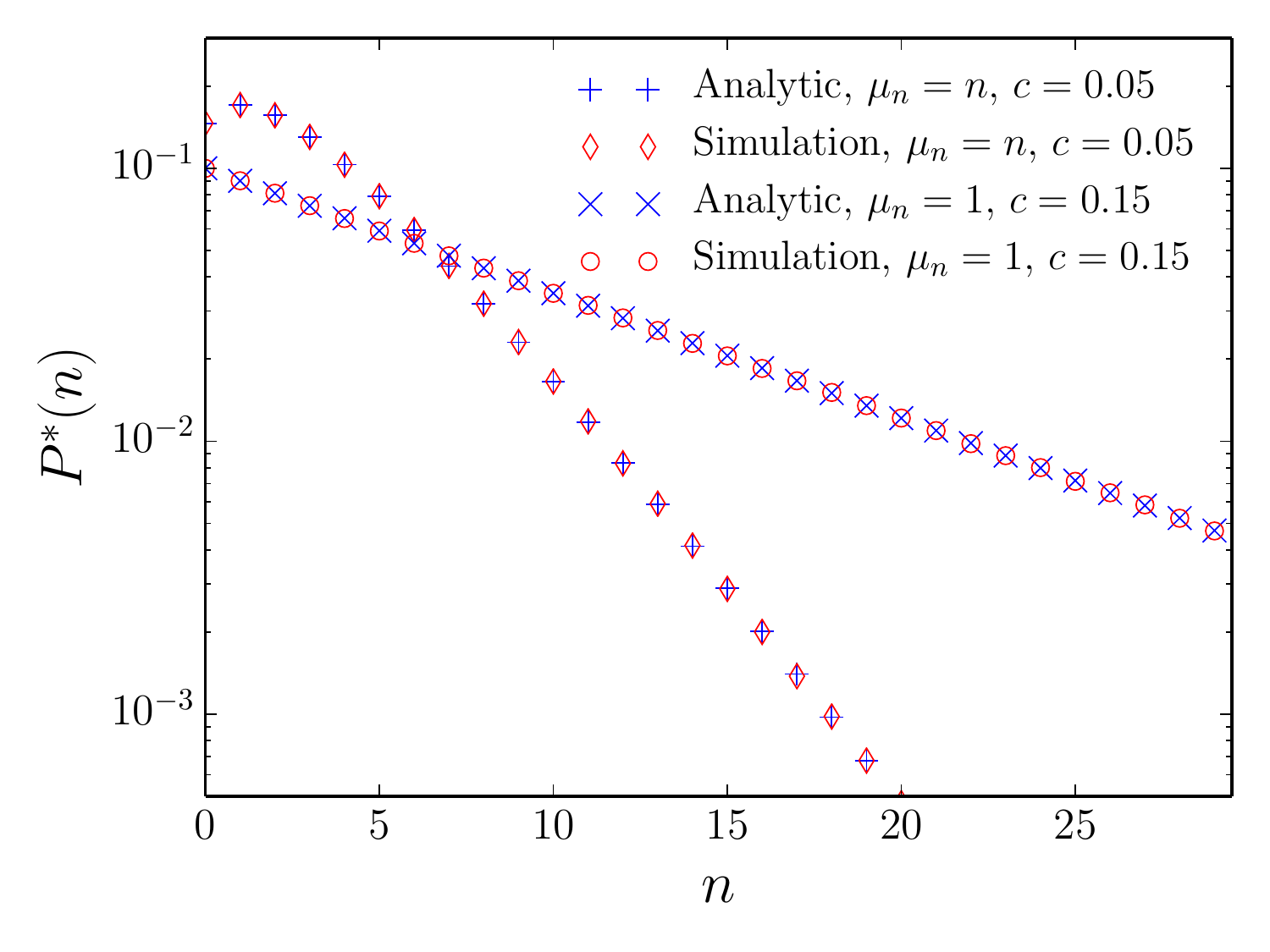}
    \caption{\label{fig:distribution} (Color online) Occupation probability distribution of the one-site system for
    constant ($\m_n=\m, n>0$) and linear ($\m_n=n$) microscopic departure rate. 
    The arrival and departure rate are $\a=0.1$ and $\b=0.2$ respectively.
    }
\end{figure}

By equating the left-hand sides of Eqs.~(\ref{eq:1})--(\ref{eq:4}) to zero, we find that the stationary distribution is given by
\begin{align}
	P^*(n) =&   \frac{z^n}{Z_c} \prod_{i=1}^{n}  w_{c,i}^{-1}, \label{eq:stationary_on_off_1}  \\
	P^*(n,\off) =& \frac{\b \m_n}{\a + c + \b \m_n}  P^*(n),\\
	P^*(n,\on) =& \frac{(\a +c )}{\a + c + \b \m_n}  P^*(n), \label{eq:stationary_on_off_3}
\end{align}
where $z=\a/\b$,
$w_{c,i} = \mu_{i}(\a + c)/(\a + c + \b \m_{i})$,
$Z_c=\sum^\infty_{n=0} z^n \prod_{i=1}^{n}  w_{c,i}^{-1}$,
and $P^*(n)=P^*(n,\on)+P^*(n,\off)$ by construction.
We recognise the same stationary state~\eqref{eq:stationary_uno} as the standard ZRP, with an effective departure rate
$w_{c,n} = \m_n P(\on|n)$, where $P(\on|n)=(\a +c )/(\a + c + \b \m_n)$ is the conditional probability of finding
the site in the \ON state, given that there are $n$ particles.
For $c \to \infty$, the effective jump rate converges
to the microscopic rate, i.e., $w_{c,n} \to \m_n$.
The stationary probability distribution of the occupation number is checked against Monte Carlo simulations in 
Fig.~\ref{fig:distribution} for both constant and linear departure rates.
Its tail is longer than the corresponding Markovian model ($c \to \infty$).
The derivation of~\eqref{eq:stationary_on_off_1}--\eqref{eq:stationary_on_off_3} is reported in Appendix~\ref{sec:stationary_on_off}.

The normalisation condition $\sum_n P^*(n) = 1$ on the probability
distribution~\eqref{eq:stationary_on_off_1} requires $\lim_{n \to \infty} \a /( \b w_{c,n}) < 1$.
For $\mu_n=\mu$, the effective departure rate is referred to as $w_c$,
and this stationarity condition is simply $ \a /( \b w_{c} ) < 1$.
It implies that values of $c$ smaller than the threshold $c_{1} := \a^2/( \b \m-\a)$
exclude any stationary state and produce a congested phase.
The onset of congestion in a larger system with constant departure rate
is explored in Sec.~\ref{sec:mean_field}  using a mean-field approach.
For unbounded microscopic departure rates, i.e., $\lim_{n \to \infty} \m_n= \infty $,
the effective interaction is still bounded, since $\lim_{n \to \infty} w_{c,n} =(\a + c)/\b$.
However, as long as $c>0$, the normalisation condition is always ensured.
Obviously, the linear departure rate case  falls into this category.

We now outline the quantum Hamiltonian representation~\cite{Schutz2001} of the Master equation~(\ref{eq:1})--(\ref{eq:4}),
which will turn out to be convenient for the study of fluctuations in Sec.~\ref{sec:fluctuations}.
The epithet ``quantum'' has become standard in the literature on interacting-particle systems,
along with the warnings that underline that the generator of a stochastic process is in general non-Hermitian, contrary to the
operators in quantum mechanics. In this approach one works in the joint occupation and phase
vector space, defining a probability basis vector
$|n,\tau \rangle  = |n\rangle \otimes |\tau\rangle $ representing a configuration with $n$
particles
and phase $\tau$.
A probability vector
$|P (t) \rangle = \sum_{n,\tau} P(n,\tau; t) |n,\tau\rangle$
obeys the normalisation condition $\langle 1|P (t)\rangle=1$
where $\langle 1| = (1,1, \ldots)$.
The Master equation then reads as
\begin{equation}
    \dfrac{d }{dt} |P(t)\rangle = - H |P(t)\rangle,
    \label{eq:master_equation}
\end{equation}
where the operator $H$ is a single-site Hamiltonian.
Our convention is to use the basis kets $(1,0)^T$ and $(0,1)^T$ for the
states $|\off\rangle$ and $|\on\rangle$, respectively.
A configuration with $n$ particles is represented by a basis ket with the $n$-th component
equal to $1$ and the remaining components equal to zero.
Consequently, the Hamiltonian is written as
\begin{equation}
H = -c  (a^{+}_{T_1} -   g_{T_1}  ) - \a ( a^{+}_{N_1}  f_{T_1} -
\mathds{1}) -  \b (a^-_{N_1} d_{T_1}  -  d_{N_1}  d_{T_1} ),
\label{eq:Hamiltonian_one_node_TA}
\end{equation}
with
\begin{equation}
\begin{aligned}
&a^{+}_{T_1} = 
	 \mathds{1}  \otimes \left( \begin{array}{cc}
		0 & 0 \\
		1 & 0 
	\end{array} \right)\hspace{-0.3em},
\hspace{0.1em}
f_{T_1} = 
	\mathds{1}  \otimes  \left( \begin{array}{cc}
		1 & 1 \\
		0 & 0 
	\end{array} \right)\hspace{-0.3em},
\\ 
&g_{T_1} =
    \mathds{1}  \otimes  \left( \begin{array}{cc}
        1 & 0 \\
        0 & 0 
        \end{array} \right)\hspace{-0.3em},
\hspace{0.1em}
d_{T_1} =
    \mathds{1}  \otimes  \left( \begin{array}{cc}
        0 & 0 \\
        0 & 1 
        \end{array} \right)\hspace{-0.3em},
\end{aligned}
\end{equation}
\begin{align}
a^{+}_{N_1} &= 
	\left( \begin{array}{cccc}
		0 & 0 & 0 & \ldots\\
		1&0&0& \\
		0&1&0&\\
		\vdots&&&\ddots
	\end{array} \right) \otimes \mathds{1}  ,
\\
a^-_{N_1} &= 
	 \left(  \begin{array}{ccccc}
		0 & \mu_1 & 0 & 0 & \ldots\\
		0&0&\mu_2&0& \\
		0&0&0&\mu_3&\\
		0&0&0&0&\\
		\vdots&&&&\ddots
	\end{array} \right) \otimes \mathds{1} ,
\\
d_{N_1}  &= 
	 \left( \begin{array}{ccccc}
		0 & 0 & 0 & 0 & \ldots\\
		0&\mu_1 &0&0& \\
		0&0&\mu_2&0&\\
		0&0&0&\mu_3&\\
		\vdots&&&&\ddots
	\end{array} \right) \otimes \mathds{1}.
\end{align}
We use the convention that a ladder operator with subscript $N_1$ or $T_1$ acts  non-trivially
only on the occupation or phase subspace respectively.
The additional subscripts $1$ underline that this Hamiltonian generates the dynamics
for the single-site case.
The operator $H$ has a block tridiagonal structure which occurs in general
in stochastic generators of processes with two variables, $n$ and $\t$ in this case.
The blocks correspond to changes in the first variable, while the entries inside
the blocks correspond to changes of the second one.
All the variables can change by at most $1$. Such processes belong to the class of
\textit{quasi-birth-death} processes and are simple cases of queues with
Markovian arrival and general departure law~\cite{Neuts1981,Stewart2009probability}.
We mention also that the results in this section can be adapted to the more general PA case
with the replacement $\a \to \a +\d$ and $\b \to \b + \g $.
Specifically, the quantum Hamiltonian for PA dynamics on a single-site is

\begin{multline}
        H = - c ( a^{+}_{T_1} -  g_{T_1})
 - \a( a^{+}_{N_1}  f_{T_1} - \mathds{1} )
 - \b( a^-_{N_1}    d_{T_1} -  d_{N_1}  d_{T_1}) \\
 - \g( a^-_{N_1}    d_{T_1} -  d_{N_1}  d_{T_1} )
 - \d( a^{+}_{N_1}  f_{T_1} - \mathds{1} ).
        \label{eq:Hamiltonian_one_node_PA}
\end{multline}

\subsection{Mean-field solution for the $L$-site system}
\label{sec:mean_field}
In the case considered so far, particles arrive on the site from the boundaries according to a Poisson process.
The many-site system is rather more complicated than this.
In fact, each site receives particles according to a more general point process, which alternates time intervals
with no events (corresponding to the \OFF phases of the neighbour sites) and periods with arrivals.
Moreover, the exact statistics of the phase switching is not \textit{a priori} known.

In this subsection, we derive an approximate solution for the stationary state of the on-off ZRP on an open chain.
The approximation consists in decoupling the equations which describe the dynamics for each site, replacing the
point process that governs the arrival on each site by a Poisson process with an effective characteristic rate.
The decoupling of the equations allows us to use the results obtained for the one-site system (Sec.~\ref{sec:stationary_distribution}).

Let us first consider the general model described in Sec.~\ref{sec:model}, where the departure rates $\mu_{n,\t}$ retain a non trivial dependence on both
$n$ and $\t$. 
We assume a product measure $\prod_{l=1}^{L} P_l^*(n_l,\tau_l )$ for the joint
probability $P^*(\{n_l,\tau_l\})$ that the system is in a steady state
with the generic site $l$ in configuration $(n_l,\t_l)$. Imposing this solution
in the stationarity condition of the $L$-site Master equation, we get
\begin{multline}
    c P_l^*(n_l, \tau_l-1) 
    + p  \sum_{\tau} z_{l-1}  P_l^*(n_l -1, \tau) \delta_{\tau_l,0} \\
    + q  \sum_{\tau} z_{l+1}  P_l^*(n_l -1, \tau) \delta_{\tau_l,0} 
    + (p + q) \mu_{n_l+1,\tau_l} P_l^*(n_l+1,\tau_l) \\
    - [ p z_{l-1} + q z_{l+1} + (p + q) \mu_{n_l,\tau_l} + c] P_l^*(n_l,\tau_l) = 0,
    \label{eq:mean_field_bulk}
\end{multline}
for the generic bulk site $l$, $1<l<L$.
We use the symbol $z_l$, already adopted in Sec.~\ref{sec:stationary_distribution} for the fugacity,
to denote the ensemble average of the departure rate,
since $z_{l} =  \sum_{n_l,\tau_l} \mu_{n_{l},\tau_{l}} P^*_l(n_l,\tau_l )$.
The use of an average interaction term justifies the appellation mean-field.
Similarly, for the leftmost and rightmost sites we get, respectively,
\begin{multline}
    c P_1^*(n_1, \tau_1-1)
    + \a  \sum_{\tau}  P_1^*(n_1 -1, \tau) \delta_{\tau_1,0} \\
    + q  \sum_{\tau} z_{2}  P_1^*(n_1 -1, \tau) \delta_{\tau_1,0} 
    + (p + \c) \mu_{n_1+1,\tau_1} P_1^*(n_1+1,\tau_1) \\
    - [ \a + q z_2 + (p + \c) \mu_{n_1,\tau_1} +c ] P_1^*(n_1,\tau_1) = 0
    \label{eq:mean_field_left}
\end{multline}
and
\begin{multline}
              c P_L^*(n_L, \tau_L-1)
            + p   \sum_{\tau} z_{L-1}  P_L(n_L - 1, \tau) \delta_{\tau_L,0} \\
            + \d  \sum_{\tau}  P_L(n_L -1, \tau) \delta_{\tau_L,0} 
            + (\b + q) \mu_{n_L+1,\tau_L} P_L^*(n_L+1,\tau_L) \\
            - [  p z_{L-1} + \d + (\b + q) \mu_{n_L,\tau_L}  +c] P_L^*(n_L, \tau_L)  = 0.
    \label{eq:mean_field_right}
\end{multline}
In Eq.~\eqref{eq:mean_field_bulk} we recognise the stationarity condition for the single site with arrival and departure
rates equal to $(p z_{l-1} + q z_{l+1} ) $ and $(p + q) \mu_{n_l,\tau_l}$, respectively. Similarly, Eq.~\eqref{eq:mean_field_left}
 is the stationarity condition for a single site with arrival and departure rates equal, respectively, to $(\a + q z_{2}) $
and $(p + \c) \mu_{n_1,\tau_1}$, while Eq.~\eqref{eq:mean_field_right} has arrival and departure rates equal, respectively, to
 $(p z_{L-1} + \d) $ and $ (\b + q) \mu_{n_L,\tau_L}$.
These conditions, in the simplified case $\mu_{n,\tau}=\m_n$ when $\tau>0$,
allow us to write an approximate stationary
distribution for each site $l$ analogous to~\eqref{eq:stationary_on_off_1} but with modified hopping rates
\begin{equation}
    P^*_l(n)=\frac{ {z_l}^n}{Z_{c,l}} \prod_{i=1}^n w^{-1}_{c,i;l}
    \label{eq:meanfield_stationary}
\end{equation}
with $Z_{c;l}=\sum_{n}{z_l}^n \prod_{i=1}^n w^{-1}_{c,i;l}$ and
\begin{equation}
\small
    \begin{aligned}
        z_{1} = &\frac{\a+ q z_{2} }{p + \c},  && w_{c,i;1} =  \frac{\mu_{i} (c + \a + q z_{2}) }{ c + \a + q z_{2} + (p+\c) \mu_{i} }, \\
        z_{L} = &\frac{\d + p z_{L-1} }{\b+q},  && w_{c,i;L} =  \frac{\mu_i (c+\d+p z_{L-1})}{c+\d+p z_{L-1}+(\b+q) \mu_{i}}, \\
        z_{l} = &\frac{p  z_{l-1} + q z_{l+1} }{p + q}, && w_{c,i;l}=   \frac{\mu_{i} (c + p z_{l-1} + q z_{l+1} ) }{ c + p z_{l-1} + q z_{l+1} +(p+q) \m_{i} },
    \end{aligned}
\label{eq:mean_field_mu_w}
\end{equation}
where $1<l<L$. The scenario of Eqs.~\eqref{eq:mean_field_mu_w} corresponds to a one-dimensional lattice where each site $l$ receives a uniform
particle stream of rate $q z_{l+1}$ ($p z_{l-1}$) from the right (left) neighbour and sends
particles according to its internal dynamics.
The consistency condition of the $z_l$ in Eq.~\eqref{eq:mean_field_mu_w},
which is equivalent to the conservation  of the current along the chain,
 is satisfied for
\begin{equation}
    \a - \c z_1 = p z_l -q z_{l+1} =  \b z_L - \d = \langle j \rangle,
    \label{eq:mf_current_conservation}
\end{equation}
$l=1,2,\ldots,L$.
The solution of this recursive relation yields the fugacity $z_l$ and the mean current $\langle j \rangle$~\cite{Levine2005}
\begin{gather}
z_l= \frac{\alpha  \beta  \left(\frac{p}{q}\right)^{L-1} -\gamma  \delta -\left(\frac{p}{q}\right)^{l-1} [\alpha  \beta -\gamma  \delta -(\alpha +\delta ) (p-q) ] }{\beta  \left(\frac{p}{q}\right)^{L-1} (\gamma +p-q) - \gamma  (\beta -p+q)}, \label{eq:fugacities}\\
\langle j \rangle =   (p - q) \frac{\a \b \left(\frac{p}{q}\right)^{L-1}  -\g \d  }{  \g(p -q -\b)+ \b(p-q+\g)      \left(\frac{p}{q}\right)^{L-1}   },
\end{gather}
which complete the mean-field solution for the model.

The approximation results in the separation of the dynamics of each site, consequently,
the mean-field quantum Hamiltonian can be written as:
\begin{equation}
    H_{\text{mf}} = H_{\text{left}} + H_{\text{right}} + \sum_{l=2}^{L-1} H_{l},
    \label{eq:mean_field_hamiltonian}
\end{equation}
where $H_{\text{left}}$, $H_{\text{right}}$, and $H_{l}$ are obtained from the generic one-site Hamiltonian
\eqref{eq:Hamiltonian_one_node_PA} using the mean-field arrival rates.

\begin{figure}
 \centering
 \includegraphics[width=8.4cm]{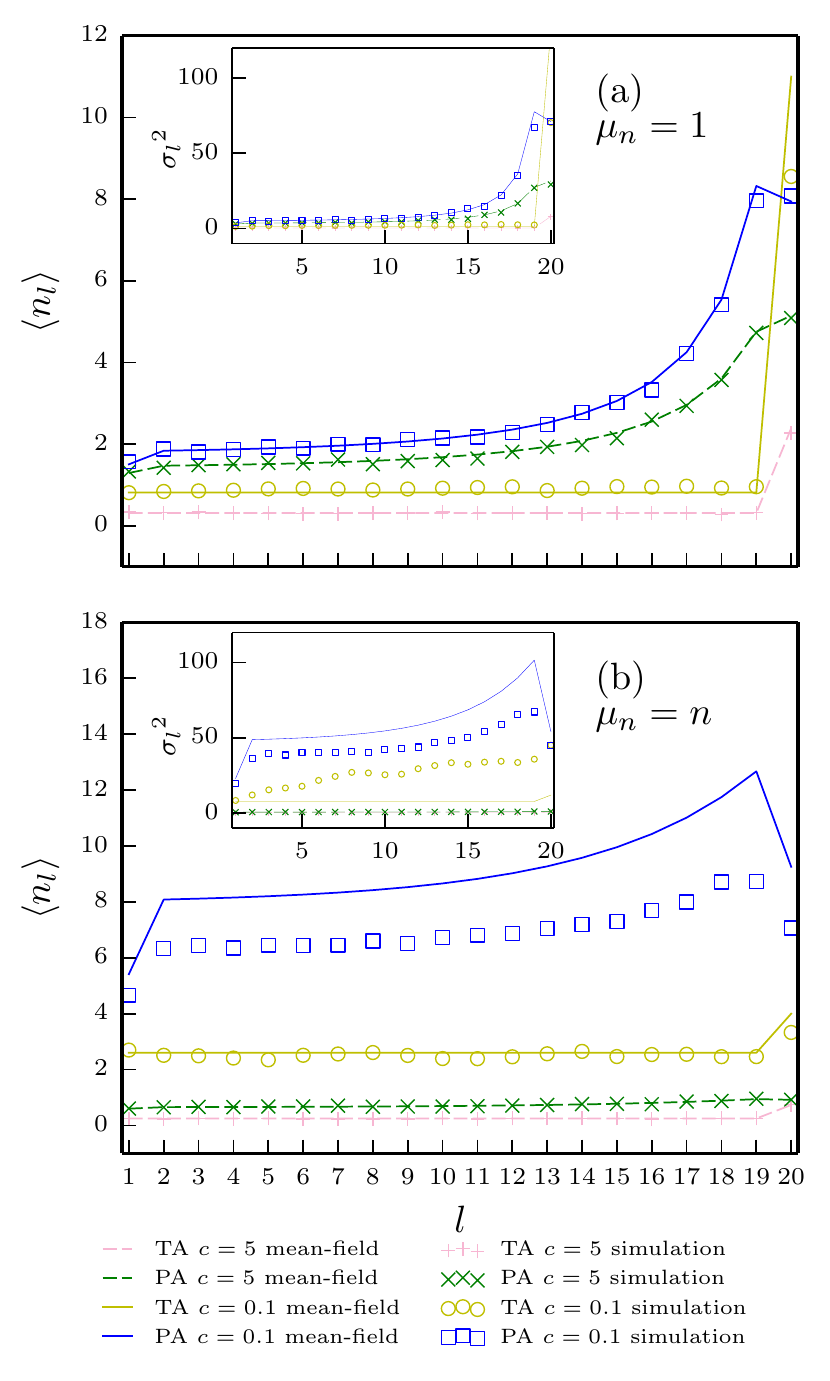}
 \caption{\label{fig:comparison}
 (Color online)
 Density profile $\langle n_l \rangle$, and variance profile  ${\sigma_l}^2$ (insets),
 on a chain of length $L=20$ with (a) constant and (b) linear departure rates.
 The results obtained within the mean-field approximation (line vertices) are compared
 with the results computed by means of Monte Carlo simulations (markers).
 TA and PA refer to rates $(\a,\b,\g,\d,p,q)=(0.2,0.3,0,0,1,0)$ and $(\a,\b,\g,\d,p,q)=(0.1,0.2,0.1,0.1,0.55,0.45)$
 respectively.
 }
\end{figure}

\begin{figure*}
 \centering
 \includegraphics[width=\textwidth]{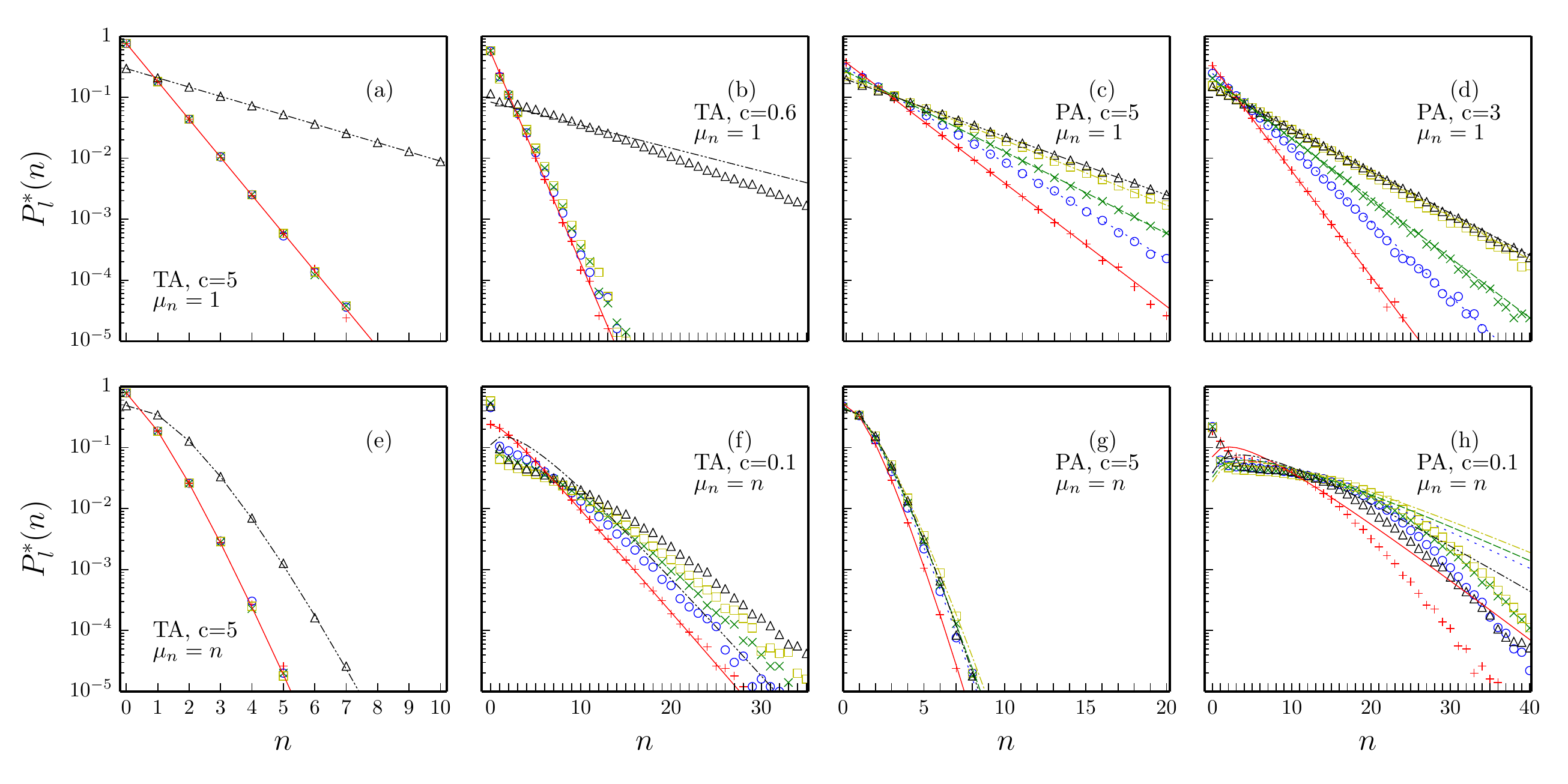}
   \caption{\label{fig:mean_field_occ_distr}
            (Color online) The mean-field probability distribution (line vertices) of the site occupation
            numbers in a chain of length $L=5$  are checked against simulations (markers).
            Symbols $+,\times,\circ,\Box,\triangle$ and solid, dotted, dashed,  dot--dashed, dot--dot--dashed lines
            of the corresponding color (grayscale) refer to sites $l=1,2,3,4,5$ respectively.
            Parameter combinations as in Fig.~\ref{fig:comparison}.
            }
\end{figure*}

\begin{figure}
    \centering
    \includegraphics[width=8cm]{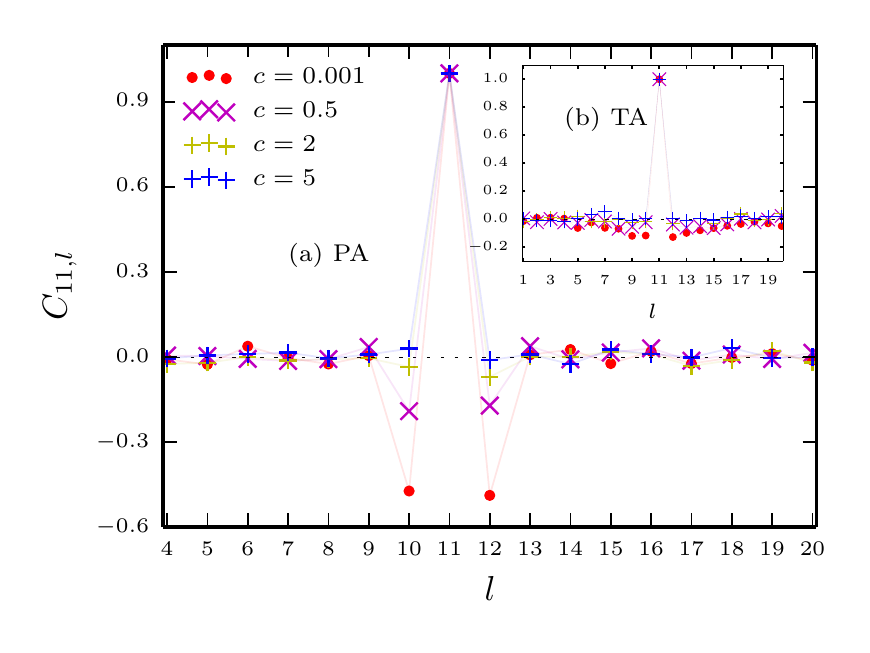}
    \caption{(Color online) Simulation results for the
    cross-correlation $C_{11,l}$ of on-off ZRP on
    a chain of length $L=20$ with $\mu_n=n$.
    (a) PA case. Adjacent sites have negatively-correlated occupation numbers.
    Hopping-rate combinations as in Fig.~\ref{fig:comparison}.
    Inset, (b). TA case. Spatial cross-correlations appear
    weaker than the PA case, but with longer range.
    As $c$ grows the correlation is gradually lost and a factorised
    solution is realistic.}
    \label{fig:cross_correlation}
\end{figure}

While the fugacities~\eqref{eq:fugacities}  are identical to those of a standard ZRP
on an open chain~\cite{Levine2005}, the effective
departure rates $w_{c,n;l}$ are affected by the time correlations
and, significantly, become site dependent.
This is evident at the level of stationary density and variance profile, respectively
$ \langle n_l \rangle = z_l \partial ( \ln Z_{c,l}) / \partial z_l $ and
$ {\sigma_l}^2 \equiv  \langle {n_l}^2 \rangle  -{\langle n_l \rangle}^2= {z_l}^2 \partial^2 ( \ln Z_{c,l}) /{\partial z_l}^2 $ in the mean-field approximation.
The predicted density profile can be non-monotonic, which contrasts with the
stationary profile of the standard ZRP~\cite{Levine2005}.
This feature is indeed present in the Monte Carlo simulated density profiles for certain parameter combinations.
In fact, for the parameters considered, the agreement between mean-field theory and simulation is excellent,
except when $c$ is very small~(see Fig.~\ref{fig:comparison}).
In Fig.~\ref{fig:mean_field_occ_distr} the mean-field predictions
for the per-site occupation distributions are checked against simulations.
The agreement is again good except for the cases with smaller values of $c$, for which the cross-correlation 
 $C_{ij}= (\langle n_i n_j\rangle - \langle n_i \rangle \langle n_j \rangle)/(\sigma_{i} \sigma_{j})$
between the occupations on site $i$ and $j$ appears to be stronger.
This is clear in Fig.~\ref{fig:cross_correlation}, where we report a negative cross-correlation between neighbouring-site occupations
for small values of~$c$.

In the cases explored above, the values of $c$ have been chosen in order to guarantee the existence of
a well defined NESS with constant average occupation number.
In the one-site system seen in Sec.~\ref{sec:stationary_distribution} this choice was straightforward,
as we can derive exactly the congestion threshold.
In an extended system with unbounded departure rates, we expect that any strictly positive value of $c$
guarantees the NESS because, although a large number of particles can pile up during the \OFF phase,
they can be released arbitrarily quickly during the \ON phase.
On the contrary, the extended system with bounded departure rates appears to be more interesting.
For values of $c$ smaller than a certain value, the particles accumulate
on one or more of the lattice sites.
We now compare the prediction of the mean-field theory for this congestion threshold
with the results of Monte Carlo simulations performed on a chain of length $L=20$.
In order to evaluate numerically the onset of congestion, we make use of the parameter (inspired by~\cite{Arenas2001})
\begin{equation}
    \kappa = \frac{n_{\text{tot}}(t+\Delta t) - n_{\text{tot}}(t)}{\Delta t} \frac{1}{ (\a+\d)},
\end{equation}
where $t \gg \Delta t$ and $n_{\text{tot}}(t) = \sum_{l=1}^L \langle n_l (t) \rangle$ is the average total number of particles in the system
at time $t$. The parameter $\kappa$ measures the difference between the rate at which
particles arrive and the rate at which particles leave the system,
scaled with respect to the total arrival rate. The congestion occurs when $\kappa$ is strictly positive~\footnote{Precisely at the threshold, we expect congestion/condensation but with sublinear growth in~time.}.
For the one-site model~\eqref{eq:1}--\eqref{eq:4} with $\mu_n=\mu$, it is straightforward to show that the expected value of $\alpha \kappa$
is the positive part of the average growth rate $ \a- \b \mu c/(c+\a) $.
This allows us to approximate a local $\kappa_l$ for the generic site $l$ of a chain,
by replacing $\a$ and $\b$ with the mean-field arrival and departure rates, respectively.
The numerical Monte Carlo study of $\kappa_l$ reveals the first site where the congestion sets in, tuning $c$ from large to smaller values.
In a chain with TA jumps, this occurs on site $1$, for $p \le \b$,
or on the site $L$, otherwise. In the PA case, the congestion can set in on the bulk site $L-1$,
as suggested by the non-monotonic density profile of Fig.~\ref{fig:comparison}.
Not surprisingly, the mean-field theory predicts this possibility.

\begin{figure}
 \centering
   \includegraphics[width=8cm]{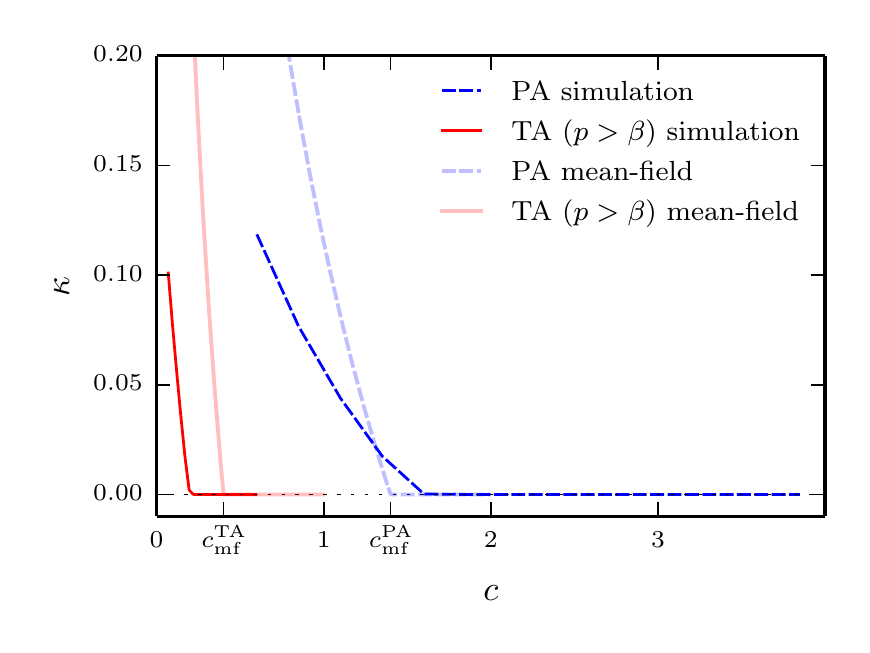}
     \caption{   \label{fig:cong_par} (Color online) Congestion transition for the on-off ZRP with $\mu=1$ on
     a open chain of length $L=20$.
     The parameter $\kappa$ obtained by Monte Carlo simulations is plotted against $c$ for a TA case and a PA case.
     Hopping-rate combinations as in Fig.~\ref{fig:comparison}.
     The mean-field congestion thresholds are $c_{\mathrm{mf}}^{\mathrm{TA}}=0.4$ and $c_{\mathrm{mf}}^{\mathrm{PA}} \simeq 1.4$ respectively. 
     These values are pinpointed by the mean-field approximated local $\kappa_l$ (light lines) of the site $l$ where the congestion sets in first.
     }
\end{figure}

We define the mean-field congestion threshold $c_{\mathrm{mf}}$ as the smallest
value of $c$ such that none of the sites $l$ of the system with Hamiltonian~\eqref{eq:mean_field_hamiltonian} has $\kappa_l>0$.
In the TA case, as long as $p>\b$, $c_{\mathrm{mf}}$ is equivalent to the
threshold $c_1$ derived in Sec.~\ref{sec:stationary_distribution} for the one-site system with boundary rates $\a$ and $\b$.
The numerical evaluation of $\kappa$ for the whole system, plotted against the mean-field estimate in Fig.~\ref{fig:cong_par}, suggests 
that $c_{\mathrm{mf}}$ is an upper bound for the true congestion transition in this case.
This relation arises as the TA jumps set the system in a highly organised
configuration, with wave-like fronts which are precursors of the slinky motion
observed in~\textcite{Hirschberg2009,Hirschberg2012} and enhance the particle transport.
When $ p < \beta$, $c_{\mathrm{mf}} = \alpha^2/(p \mu - \alpha)$ marks exactly
the onset of the congested phase.
Conversely, PA interactions seems to promote congestion, as there are many jumps which
block the site and contribute negatively to the particle current. In this case the congestion
transition occurs for a value of $c$ larger than both $c_1$ and $c_{\mathrm{mf}}$~(see Fig.~\ref{fig:cong_par}).

In the next section the out-of-equilibrium aspects of this model are
further investigated by focusing on the fluctuations of the particle currents.

\section{Current fluctuations}
\label{sec:fluctuations}
This section is devoted to the study of the full statistics of the empirical currents
$j_l=J_l/t$, where $J_l$ is the difference between the number of particle hops
from site $l$ to site $l+1$ and the number of hops from site $l+1$ to site $l$.
This definition is extended to the input current $j_0$
and to the output current $j_L$.
In order to lighten the notation, we simply make use of $j$ and $J$ and explicitly specify
the bond only when necessary.
For $t \to \infty$, $j$ converges to its ensemble average $\langle j \rangle$.
However, for finite time it is still possible to observe fluctuations of $j$
from this typical value. These fluctuations are quantified by means of
the \textit{scaled cumulant generating function}
 $e(s)$ or the \textit{rate function} $\hat{e}(j)$.
In the following we define these concepts.

\subsection{Large deviation formalism}
\label{sec:large_deviations}
Generically, in the long-time limit, the probability $p(j,t)$ of observing a current
$j$ at time $t$ obeys a large deviation principle of the form
\begin{equation}
    p(j,t) \sim e^{-t \hat{e}(j)}.
    \label{eq:large_deviation_principle}
\end{equation}
To obtain the rate function $\hat{e}(j)$,
we first investigate the moment generating function of the total integrated current $J$:
\begin{equation}
    \langle  e^{-s J} \rangle  = \langle 1 | e^{-s \hat{J}} | P_J(t) \rangle,
    \label{eq:moment_generating_function}
\end{equation}
where $\hat{J}$ is a diagonal operator whose diagonal elements are the set of possible values that the integrated current can assume
and the probability distribution $| P_J (t)  \rangle = \sum_{n,\tau,J} P(n,\tau,J;t) |n,\tau,J \rangle$ is now defined in the occupation,
clock and current configuration space.
The distribution $| P_J (t)  \rangle $ can be obtained from a generic initial state $| P_0 (0) \rangle$ with $(t,J)=(0,0)$  through
$| P(t)  \rangle = e^{-H_J t} | P_0 (0) \rangle $,
where $e^{-H_J t}$ is the time evolution operator in the joint configuration and current space.
We diagonalise the operator $e^{-H_J t}$ by means of a Laplace transform
$e^{-s \hat{J} } e^{-H_J t} e^{s \hat{J}}$. In the configuration subspace, this reduces to $e^{-\tilde{H} t}$,
where the operator $\tilde{H}$ is obtained  multiplying  by $e^{-s}$ (or $e^{s}$)
the entries of the original Hamiltonian $H$ which produce a unit increase (or decrease) in $J$~\cite{Harris2005}.
Hereafter, we refer to the tilded operator $\tilde{H}$ as the $s$-modified Hamiltonian.
Since  $\langle 1| e^{-s \hat{J} } | P_{0} (0) \rangle=1$,
then $\langle  e^{-s J } \rangle = \langle 1| e^{-\tilde{H} t} | P(0)\rangle$,
where $| P(0)\rangle $ now denotes a probability vector in the subspace of the occupation number and the clock variable.
Let us denote by $|\til{P}_{A_0} \rangle$ the right eigenvector of $\til{H}$ associated with the discrete smallest eigenvalue $A_0$.
The long-time limit of the generating function is accessible through
\begin{equation}
 \langle  e^{-s J} \rangle \sim   \langle 1|  \til{P}_{A_0}\rangle  \langle \til{P}_{A_0}|   P(0) \rangle e^{-A_0 t}, \qquad   t \to \infty,
 \label{eq:naive}
\end{equation}
as long as the pre-factors $\langle 1| \til{P}_{A_0}\rangle $ and $\langle \til{P}_{A_0}|  P(0)\rangle$
are finite and a point spectrum exists [see, e.g.,~\cite{Harris2005}].

Although the moment generating function and the conjugated variable $s$ have an analogue in equilibrium statistical mechanics, i.e.,
the Helmholtz free energy and the pressure, respectively, they are not as readily  accessible
(we cannot tune $s$ as we can do with the pressure or temperature).
However, the generating function~\eqref{eq:moment_generating_function} helps to find out the rate function.
In fact, as long the limit relation~\eqref{eq:naive} is valid,
we can identify $A_0$ with the scaled cumulant generating function (SCGF)
\begin{equation}
    e(s) =  - \lim_{t \to \infty} \frac{1}{t} \ln \langle  e^{-s J } \rangle.
    \label{eq:cumulant}
\end{equation}
The SCGF in turn gives the convex hull of the rate function though a Legendre-Fenchel transform~\cite{Touchette2009}:
\begin{equation}
    \hat{e}(j) =  \text{sup}_s \{ e(s) - s j\}.
    \label{eq:legendre}
\end{equation}
When one of the two pre-factors in Eq.~\eqref{eq:naive} diverges,
or when the spectrum is entirely continuous,
we need to employ other methods~(see Sec.~\ref{sec:large_current_fluctuations}).

\subsection{Analytical results for the single-site system}
For the single-site system, we study the fluctuations of the output current, simply denoted by $j$.
The input current can be obtained in the PA case by reflection, while in the TA case it
is given by a simple Poisson process.
Despite its simplicity, the single-site ZRP exhibits a rich fluctuating behaviour,
even in the absence of time correlations~\cite{Harris2006,Rakos2008,Harris2013}.
The introduction of the on-off mechanism  creates a still more interesting scenario.
In fact, the study of the fluctuations reveals some aspects of the correlations
which, in the stationary state, are hidden within an effective interaction factor.

\subsubsection{Small current fluctuations}
\label{sec:small_fluctuations}
The $s$-modified Hamiltonian corresponding to the output current is obtained from~\eqref{eq:Hamiltonian_one_node_PA}
multiplying  the ladder operators $\b a^-_{N_1}$ and $\d a^{+}_{N_1}$ by $e^{-s}$ and $e^{+s}$, respectively:
\begin{multline}
    \til{H} = - c ( a^{+}_{T_1} -  g_{T_1} )
  -  \a (a^{+}_{N_1}  f_{T_1} - \mathds{1} )
  -  \b (e^{-s} a^-_{N_1}  d_{T_1} -  d_{N_1}  d_{T_1})\\
  -  \g (a^-_{N_1}  d_{T_1} -  d_{N_1}  d_{T_1} )
  -  \d (e^{s} a^{+}_{N_1} f_{T_1}   - \mathds{1} ).
\label{eq:biased_hamiltonian}
\end{multline}
We concentrate now on the eigenproblem
\begin{equation}
    ( \tilde{H} - A \mathds{1}) | \tilde{P}_A \rangle =0,
    \label{eq:eigenproblem}
\end{equation}
where $| \tilde{P}_A \rangle $ is the generic right eigenvector and  $A$ is its eigenvalue.
It is convenient to write the eigenvector $|\tilde{P}_{A_0}\rangle$, associated to $A_0$,
in a form similar to the stationary solution~\eqref{eq:stationary_on_off_1}--\eqref{eq:stationary_on_off_3}, i.e., with components:
\begin{align}
  \tilde{P}_{A_0}(n,\on) &= p_{\mbox{\tiny ON},n,s} \tilde{P}_{A_0}(n), \label{eq:old_ansatz1} \\
  \tilde{P}_{A_0}(n,\off) &= (1 - p_{\mbox{\tiny ON},n,s})  \tilde{P}_{A_0}(n), \label{eq:old_ansatz2} \\
  \tilde{P}_{A_0}(n+1) &= \rho_{n+1,s} \tilde{P}_{A_0}(n). \label{eq:old_ansatz3}
\end{align}
Equation~\eqref{eq:eigenproblem} is hard to solve in general. To gain insight into the
appropriate structure of $\rho_{n,s}$  and $p_{\mbox{\tiny ON},n,s}$, 
we study first the simple case with constant departure rates.

\noindent \textbf{Constant departure rates.}
Let the  departure rate be  $\mu_n=\mu$ when $n>0$.
Motivated by the stationary state result, we assume here that the factors
$p_{\mbox{\tiny ON},n,s}$ and $\rho_{n,s}$ have no dependence on the occupation number and we drop the subscript $n$
with the exception of $n=0$,  i.e., $p_{\mbox{\tiny ON},0,s}$ is distinct from  $p_{\mbox{\tiny ON},s}$.
By direct substitution  into Eq.~\eqref{eq:eigenproblem} we get:
\begin{align}
        -(c + \alpha  + \delta  - A_0) (1 - p_{\mbox{\tiny ON},0,s}) = 0,  \label{eq:row_equation_1} \\
   \begin{multlined}
        c (1 - p_{\mbox{\tiny ON},0,s}) - (\alpha + \delta  - A_0 ) p_{\mbox{\tiny ON},0,s} \\
        + (\beta e^{-s} + \gamma) \mu  p_{\mbox{\tiny ON},s} \rho_s  = 0,
   \end{multlined}&  \\
   \begin{multlined}
        (\alpha  + \delta e^s )  - (c + \alpha + \delta - A_0) (1 - p_{\mbox{\tiny ON},s}) \rho_s=0,
   \end{multlined} & \\
   \begin{multlined}
        c (1 - p_{\mbox{\tiny ON},s})- [\alpha + \delta +  (\beta + \gamma) \mu - A_0 ] p_{\mbox{\tiny ON},s}  \\ 
        + (\beta e^{-s} + \gamma ) \mu p_{\mbox{\tiny ON},s} \rho_s = 0.
   \end{multlined}&
\end{align}
Equation~\eqref{eq:row_equation_1} trivially requires $p_{\text{\tiny ON},s,0}=1$, while we expect that $p_{\mbox{\tiny ON},s}<1$.
After a long but straightforward algebraic manipulation, the system is solved for
\begin{align}
 p_{\mbox{\tiny ON},s} & = 
\frac{ c + (\b e^{-s} + \g)  {(\a + \d e^{s})}/{(\b+\g)} }{c + (\b+\g)\mu + (\b e^{-s} + \g)  {(\a + \d e^{s})}/{(\b+\g) } }  \\
        \rho_s &= \frac{(\alpha +\delta  e^s)}{(\beta +\gamma )} (\mu p_{\text{\tiny ON},s})^{-1},\\
        A_0 &= \frac{\a \b}{\b+\c}(1-e^{-s}) + \frac{\c \d}{\b + \c} (1-e^{s}). \label{eq:lowest_ev}
\end{align}
Note that setting $s = 0$, the factor $p_{\mbox{\tiny ON},s}$
becomes the conditional probability $P^*(\on|n)$ in the steady state.
Also, the parameter $\rho_s$ and the eigenvalue $A_0$ have a counterpart
in the stationary probability, in fact for $s \to 0$, $\rho_s \to z w_{c}^{-1}$, and $A_0 \to 0$.
Consequently, we argue that $A_0$ is the lowest eigenvalue of $\tilde{H}$ and,
according to Sec.~\ref{sec:large_deviations}, the SCGF at least in the neighbourhood of $s=0$.

For later convenience, we define a modified fugacity
\begin{equation}
    z_s = \frac{\a + \d e ^s}{\b+\g}.
    \label{eq:mod_fugacity}
\end{equation}
and a modified effective interaction
\begin{equation}
    w_{c,s}  = \m  p_{\mbox{\tiny ON},s}
    \label{eq:w_s}
\end{equation}
such that $\rho _{s}=z_s w_{c,s}^{-1}$ and $w_{c,s}\to \mu$ for $c \to \infty$.
It is worth noting that,  while the bias affects  only the fugacity in the ordinary ZRP~\cite{Harris2005},
it affects both the interaction term and the fugacity in the on-off model.

\noindent \textbf{General departure rates.}
This paragraph covers also the special case with linear departure rates $\mu_n=n$.
Motivated by the results above, we assume that
the components of the ground state eigenvector $|\til{P}_{A_0} \rangle$
satisfy Eqs.~\eqref{eq:old_ansatz1}--\eqref{eq:old_ansatz3} with
\begin{align}
    \rho_{n,s} &= z_s w_{c,n,s}^{-1},\label{eq:ansatz_general1}\\
    w_{c,n,s} &= \mu_n p_{\mbox{\tiny ON},n,s}, \label{eq:ansatz_general2}
\end{align}
for $n \ge 0$. With this assumption, the second row equation of the eigenproblem~\eqref{eq:eigenproblem} is solved for
$A=A_0 \equiv \a + \b - (\b e^{-s} + \g) z_s$ and the remaining equations yield a solution for $z_s$ consistent with~\eqref{eq:mod_fugacity} and
an $n$-dependent effective interaction
\begin{equation}
w_{c,n,s} = \frac{\m_n [c + (\b e^{-s} +\g)(\a+\d e^{s})/(\b +\g) ] }{(\b e^{-s} +\g)(\a+\d e^{s})/(\b +\g) + c + (\b+\g)\m_n}.
\label{eq:wn}
\end{equation}

The eigenvalue we obtained is the same as the lowest eigenvalue $A_0$~\eqref{eq:lowest_ev}
of the $s$-modified Hamiltonian for the standard ZRP~\cite{Harris2005}.
In fact, the affinity between the two models appears closer if 
we work in the reduced state space obtained by collapsing the states corresponding to
$\t=\on$ and $\t=\off$, for each occupation number, and considering the sum of
their non-conserved probabilities $\til{P}_{A_0}(n) = \til{P}_{A_0}(n,\on) + \til{P}_{A_0}(n,\off)$.
We notice that the vector $| \til{P}^{\star}_{A_0} \rangle$ with components $\til{P}_{A_0}(n)$ is the right eigenvector
with eigenvalue $A_0$ of
\begin{equation}
    \til{H}^{\star} = \a(a^+ -1) + \d(e^s a^+-1) + \g (a^{\star-}_s -d^{\star}_s) +\b(e^{-s} a^{\star-}_s - d^{\star}_s) ,
    \label{eq:operator}
\end{equation}
where
\begin{align}
a_s^{\star-} &= \left( \begin{array}{ccccc}
		0 & w_{c,1,s} & 0 & 0 & \ldots\\
		0&0& w_{c,2,s}&0& \\
		0&0&0& w_{c,3,s}&\\
		0&0&0&0&\\
		\vdots&&&&\ddots
	\end{array} \right),
\end{align}
\begin{equation}
a^+ = 
	\left( \begin{array}{cccc}
		0 & 0 & 0 & \ldots\\
		1&0&0& \\
		0&1&0&\\
		\vdots&&&\ddots
	\end{array} \right),
\end{equation}
and the operator ${d^{\star}_s}$ has entries $\d_{ij} w_{c,i,s}$.
The operator $\til{H}^{\star}$ is equivalent to the $s$-modified Hamiltonian of a standard ZRP
with departure rates $w_{c,n,s}$.
However, it is not a genuine $s$-modified Hamiltonian for the on-off ZRP
as it shares only the lowest eigenvalue $A_0$ with $\til{H}$ (the higher eigenvalues being different, in general)
hence it only contains information about the limiting behaviour and does not generate the dynamics.

As a partial conclusion, we underline that both the systems with bounded and unbounded rates
display the fluctuating behaviour seen in the standard ZRP as long as the ground state
satisfies Eqs.~\eqref{eq:ansatz_general1} and~\eqref{eq:ansatz_general2}.
This is certainly true for current fluctuations close to the mean $\langle j \rangle$.
However, the effective interaction $w_{c,n,s}$ has a dependence on $n$ and
$s$ different from the standard ZRP and this alters the range of validity of this regime.
In the following, we show that larger current fluctuations in the on-off ZRP
can be strongly affected by time correlations.

\subsubsection{Range of validity}
\label{sec:range_of_validity}
The scenario seen so far is an analytical continuation of the stationary state.
Despite  this, certain values of the bias $s$ correspond to 
non-analyticity in the SCGF.
Such a behaviour is often referred to as a dynamical phase transition  because of
the analogy of the SCGF with the Helmholtz free energy.
According to Sec.~\ref{sec:large_deviations}, a transition occurs
as soon as the scalar product $\langle 1 | \til{P}_{A_0} \rangle$ or
$\langle \til{P}_{A_0} | P(0) \rangle$ diverges.
The choice of the initial distribution $| P(0) \rangle$ influences the
value of the second norm.
In order to ensure a finite $\langle \til{P}_{A_0} | P(0) \rangle$,
we will always consider an empty site as initial condition, unless explicitly stated otherwise.
We must also ensure that the norm $\langle \til{P}_{A_0} | \til{P}_{A_0} \rangle$ is finite,
i.e., that the eigenvector is normalizable and the discrete eigenvalue $A_0$ exists.
We now derive exactly the conditions 
under which the norms $\langle 1 | \til{P}_{A_0} \rangle$  and $\langle \til{P}_{A_0} | \til{P}_{A_0} \rangle$ 
converge and it is possible to identify  the SCGF with the lowest eigenvalue $A_0$ given in Eq.~\eqref{eq:lowest_ev}.

\noindent \textbf{Linear departure rates.} We focus first on the case with $\m_n= n$.
For this particular choice of the interaction, 
particles in the memoryless ZRP can never pile up and the current shows a smooth SCGF.
On the contrary, in the on-off model, the particle blockade alters the statistics of
small currents.
From a mathematical point of view,  a transition occurs when $\langle 1| \tilde{P}_{A_0} \rangle$ diverges.
The condition $\langle 1| \tilde{P}_{A_0} \rangle< \infty$ is satisfied for
$\lim_{n \to \infty }\rho_{n,s} < 1 $ where $\rho_{n,s}$ is defined in Eq.~\eqref{eq:ansatz_general1}.
For later convenience, we simplify this condition as
\begin{equation}
    A_0 < c + \d(1-e^s),
    \label{eq:condition_s1_ind}
\end{equation}
which is satisfied for $s>s_1$, where
\begin{equation}
 \small e^{s_1} = 
  \frac{2 \alpha  \beta}{\alpha  \beta -\beta  \delta -\b c -c \g +\sqrt{4 \alpha  \beta ^2 \delta +(\alpha  \beta -\beta  \delta - \beta c-c \gamma )^2}} .
 \label{eq:scrit2_PAind}
\end{equation}
In the PA case, $s_1$ is always finite.
In the TA case, i.e., $(\g,\d)=(0,0)$, the critical value $s_1 = \ln [\a/(\a -c)]$ is well defined only for $c <  \alpha$.

We can  prove that, when $\mu_n=n$, the norm $\langle \til{P}_{A_0} | \til{P}_{A_0} \rangle$ is always finite.
In Appendix~\ref{sec:left_eigen_ind}, the eigenvector $\langle \tilde{P}_{A_0} |$ is derived.
Its components have a form similar to Eqs.~\eqref{eq:old_ansatz1}--\eqref{eq:old_ansatz3},
with the factors $p_{\text{\tiny ON},s}$ and $\rho_s$ replaced by $p^{\text{\tiny left}}_{\text{\tiny ON},s} \equiv 1/2$
and $\rho^{\text{\tiny left}}_s \equiv ( \beta  e^{-s}+ \gamma)/(\beta +\gamma )$ respectively. 
The series $\langle \til{P}_{A_0} | \til{P}_{A_0} \rangle$ is simplified by
summing first the pairs corresponding to the same occupation number and the
condition for convergence can be written as $\lim_{n \to \infty} {\rho^{\text{\tiny left}}_s} \rho_{s,n} < 1$, which is always satisfied.
Consequently, for linear departure rates, the only mechanism
responsible for dynamical phase transitions is the on-off clockwork, which becomes dominant
 when  $\langle 1 | \tilde{P}_{A_0} \rangle$ diverges.

\noindent \textbf{Constant departure rates.} Let us consider the case $\mu_n=1$, $n>0$.
The scalar product $\langle 1 | \tilde{P}_{A_0} \rangle$ is finite
when the $n$-independent parameter $\rho_s$ is less than 1 and a dynamical phase transition occurs at $\rho_s = 1$.
In the PA case, the solution of this equation  for $s$ involves a cumbersome cubic and therefore is not reported here.
However, in the TA case, $s_1 = \ln[\a(\a - \m \b)/(c \m \b - \a \m \b - c \a)]$.
In order to check whether $\langle \tilde{P}_{A_0} | \tilde{P}_{A_0} \rangle$
is finite, we again need the eigenvector  $\langle \tilde{P}_{A_0} |$.
As the dependence on $\mu_n$ cancels in the left eigenproblem, $\langle \tilde{P}_{A_0} |$ is the same as
the linear departure rate case, see Appendix~\ref{sec:left_eigen_ind}. 
The condition for convergence is ${\rho^{\text{\tiny left}}_s} \rho_{s} < 1$ and
the value of $s$ such that  ${\rho^{\text{\tiny left}}_s} \rho_{s} =1$ is referred to as  $s_2$.
Also here, we only report explicitly the critical bias $s_2 = - \ln[(-c + \sqrt{c^2 + 4c\b})/(2 \a)]$ for the TA case.
The values $s=s_1$ and $s=s_2$ mark the onsets of new phases.

We notice that the scenario seen so far is entirely encoded into the operator $\til{H}^\star$~\eqref{eq:operator}.
In fact, this operator not only has lowest eigenvalue $A_0$, as seen in Sec.~\ref{sec:small_fluctuations},
but the normalisation of its ground state eigenvector yields sums $\langle 1 | \til{P}^\star_{A_0}\rangle$ and
$\langle \til{P}^\star_{A_0} | \til{P}^\star_{A_0}\rangle$ that diverge at the same critical points $s_1$ and $s_2$, respectively.
In the following, we focus on the large-fluctuation regimes $s>s_1$ and $s<s_2$.

\subsubsection{Large current fluctuations}
\label{sec:large_current_fluctuations}
We employ different approaches to study the large fluctuation
regimes in the linear and constant departure rate cases.\\
\noindent \textbf{Linear departure rates.}
For this special case, we consider first a finite-capacity version of the on-off ZRP.
In fact, the SCGF on a discrete finite configuration space is always given by the smallest eigenvalue of
the $s$-modified Hamiltonian, as the prefactors in~\eqref{eq:naive} are always finite.
For the TA case, we truncate the Hamiltonian~\eqref{eq:Hamiltonian_one_node_TA} by imposing a reflective boundary in the state with occupation number $N$.
The resulting matrix in block form is
\begin{align}
&H_{N} \notag \\
&=\left( \begin{array}{cc|cc|cc|c|cc}
		\mathsmaller{c+\a}      &\mathsmaller{0}     &\mathsmaller{0}   & \mathsmaller{0}              &\mathsmaller{0}&\mathsmaller{0} &\dots&   &             \\
        \mathsmaller{-c}        &\mathsmaller{\a}    &\mathsmaller{0}   & \mathsmaller{-\b \m_1 }      &\mathsmaller{0}&\mathsmaller{0} &     &   &             \\
\cline{1-9}
  		\mathsmaller{-\a}       &\mathsmaller{-\a}   &\mathsmaller{c+\a}& \mathsmaller{0}              &\mathsmaller{0}&\mathsmaller{0} &     &   &             \\
        \mathsmaller{0}         &\mathsmaller{0}     &\mathsmaller{-c}       &\mathsmaller{\a+ \b \m_1}&\mathsmaller{0}&\mathsmaller{-\b \m_2 }&\dots&   &             \\
\cline{1-9}
        \vdots     &         &      &                &  &\vdots          & \ddots    &                &  \vdots                      \\
                   &         &      &                &  &                &     &\mathsmaller{0} &\mathsmaller{-\b \m_N } \\
\cline{1-9}
		           &         &      &                &  &                &     &\mathsmaller{c} &\mathsmaller{0}               \\
		           &         &      &                &  &                &\dots&\mathsmaller{-c}&\mathsmaller{\b \m_N}      \\
	\end{array} \right),
	\label{eq:truncated_hamiltonian}
\end{align}
which defines a Master equation where the $n$-th block row specifies the dynamics of the configuration
with occupation number $n$ and within each block the first (second) row corresponds to an \OFF (\on) phase.

\begin{figure}
 \centering
  \includegraphics[width=8cm]{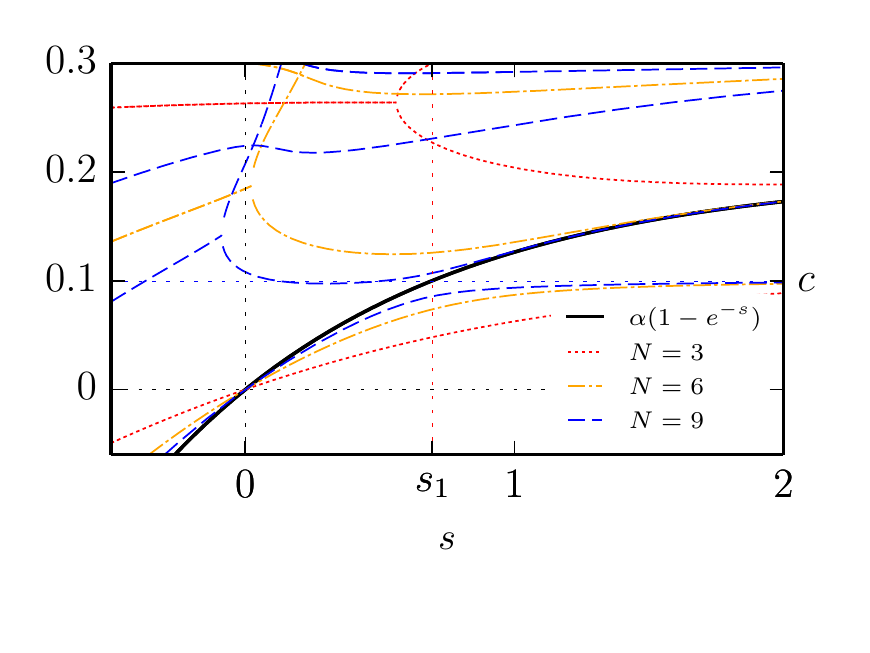}
  \caption{ \label{fig:spectr_ind} (Color online) Real part of the spectrum of the finite-capacity $s$-modified Hamiltonian for parameters
        $(\a,\b,\g,\d,c)=(0.2,0.3,0,0,0.5)$ and $\mu_n=n$.
        For $s< s_1$ the smaller eigenvalue converges to $A_0 = \a(1-e^{-s})$, while
        for $s> s_1$ it converges to $c$.
        }
\end{figure}

In the present linear departure rate case $\mu_n=n$ and the matrix $H_{N}$ generates the dynamics of a generalised exclusion process~\cite{Kipnis1999} with on-off mechanism,
or a queue with Markovian arrival times, general service and finite capacity $N$~\cite{Neuts1981,Stewart2009probability}.
According to the procedure of Sec.~\ref{sec:large_deviations}, the finite-capacity $s$-modified Hamiltonian $\tilde{H}_{N}$
is obtained by multiplying the upper-diagonal rates $\m_n$ ($n=1,2,3,\ldots,N$) of $H_N$ by $e^{-s}$.
The numerical evaluation of the spectrum of $\tilde{H}_{N}$, see Fig.~\ref{fig:spectr_ind},
shows that the two lowest eigenvalues get closer with increasing values of $N$.
This gives a clue about the limiting behaviour for $N \to \infty$,
where the eigenvalues coalesce at $s=s_1$ and two different dynamical phases emerge.
The SCGF converges to a constant branch for $s>s_1$. 
In the limit $s \to \infty$, the  truncated $s$-modified Hamiltonian is lower-diagonal
and its eigenvalues are given by the escape rates.
As long as the condition $c < \a$ holds,
the smallest eigenvalue is $c$. It corresponds  to the escape rate from the configuration with $N$ particles and \OFF state.
We expect that the corresponding eigenvector does not satisfy the ansatz~\eqref{eq:ansatz_general1}--\eqref{eq:ansatz_general2}.
Dynamical phase transitions due to the crossover of eigenvectors are observed in spatially-extended non-equilibrium models
such as the Glauber model with open boundaries~\cite{Masharian2014}.
We argue that, in the infinite capacity limit, the SCGF is given, for $s>s_1$,
by the escape rate of the system with an instantaneous congested state and \OFF state. 

\begin{figure}
 \centering
 \includegraphics[width=8cm]{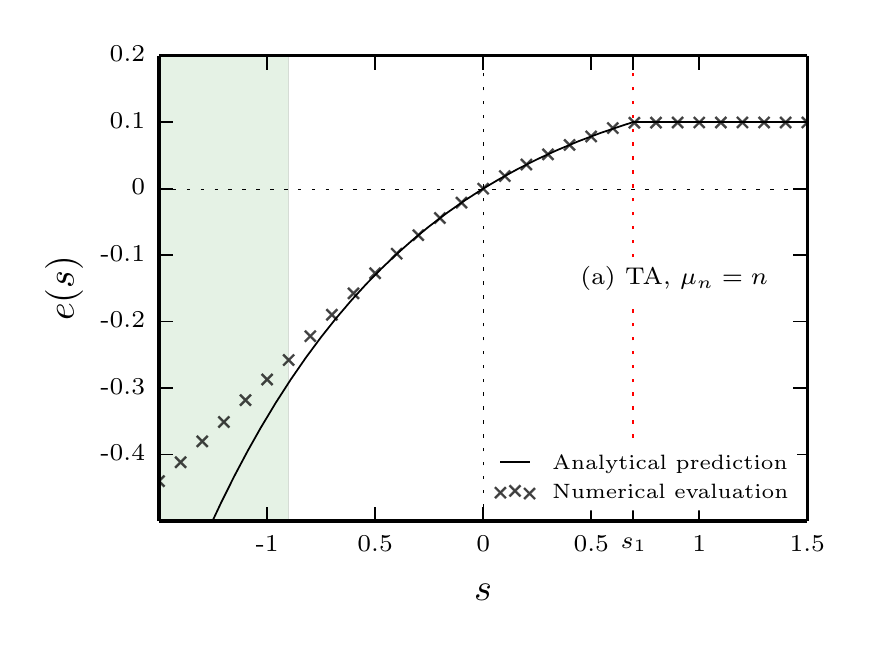}
 \includegraphics[width=8cm]{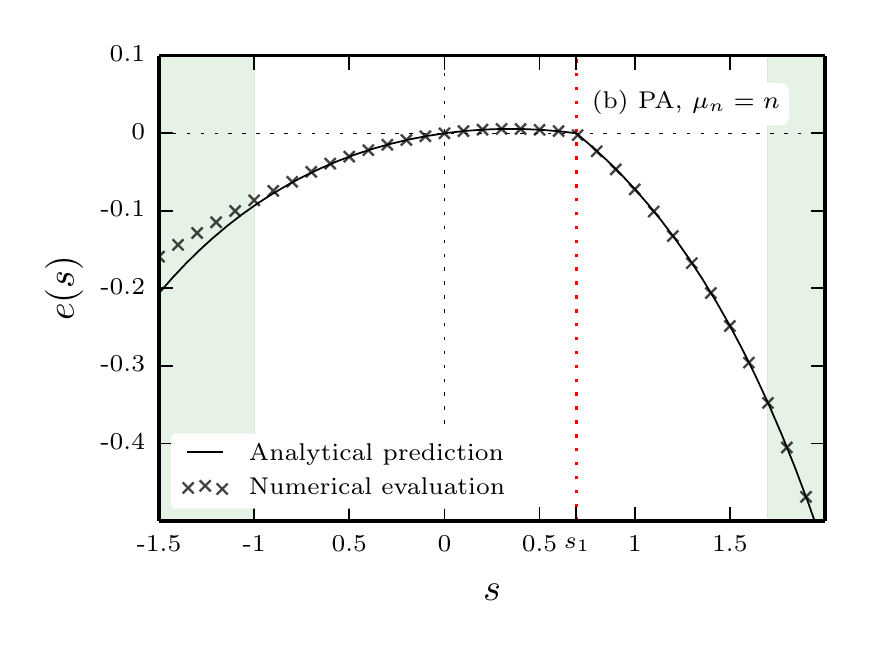}
 \caption{ \label{fig:SCGF_PA_ind} (Color online) SCGF of the on-off ZRP with $\m_n=n$ for (a) TA hopping rates, $(\a,\b,\g,\d,c)=(0.2,0.3,0,0,0.1)$, and (b)
 PA hopping rates, $(\a,\b,\g,\d,c)=(0.1,0.2,0.1,0.1,0.1)$.
 Points are data from the cloning algorithm, $\mathcal{N}=10^4,t=10^4$.
 The SCGF is systematically overestimated for small values of $s$.
 We expect a better approximation  but a slow convergence for larger ensembles and longer simulation times.}
\end{figure}

\begin{figure}
 \centering
 \includegraphics[width=8cm]{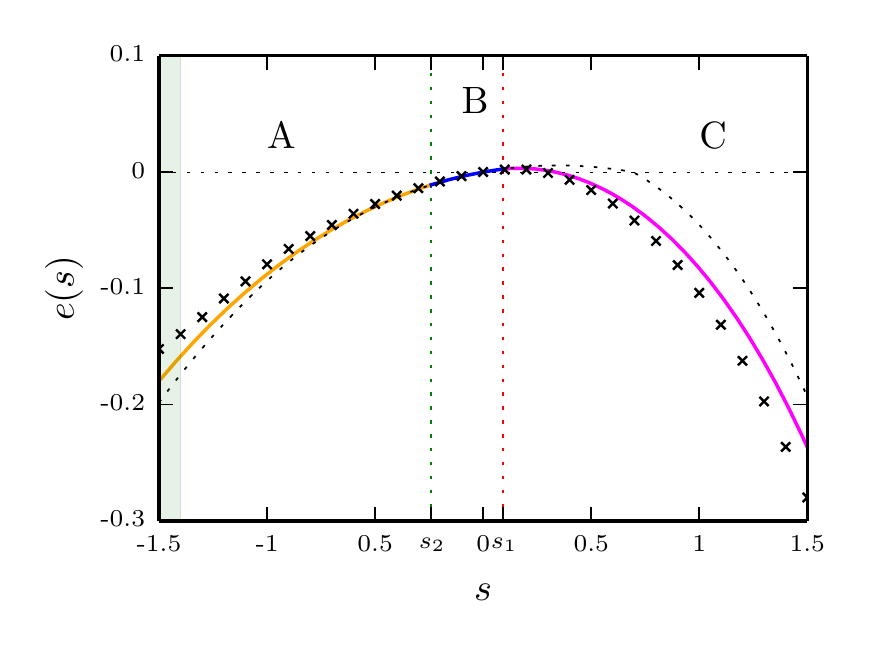}
 \caption{\label{fig:SCGF_PA_const} (Color online) SCGF of the on-off ZRP with $(\a,\b,\g,\d,c)=(0.1,0.2,0.1,0.1,0.5)$ and $\m_n=1$.
 Points are data from the cloning simulations, $\mathcal{N}=10^4, t=10^4$.
 Dotted line is the SCGF of the Markovian-ZRP ($c \to \infty$) with same boundary rates.
 Solid line is the analytic approximation~\eqref{eq:phaseA}--\eqref{eq:phaseC}.
 The SCGF of the ZRP with $s$-independent departure rate $w_{c}$
 would overlap the solid line at this scale.}
\end{figure}

Our prediction is checked against numerical simulations, as shown in Fig.~\ref{fig:SCGF_PA_ind}(a).
The simulations employ an advanced Monte Carlo algorithm, referred to as the
``cloning'' method, which allows us to measure directly the SCGF~\cite{Giardina2006, Lecomte2007}.
This method permits the integration of the dynamics generated by an $s$-modified Hamiltonian $\tilde{H}$,
by means of the parallel simulation of $\mathcal{N}$ copies of the system.
A system in state $i$ may be cloned or pruned with exponential rate $\tilde{H}_{ii} - \sum_j \tilde{H}_{ij}$,
in order to account for the fact that $\tilde{H}$ does not conserve the total probability.
The average cloning factor gives the SCGF.
This prescription is believed to be exact for $\mathcal{N} \to \infty$, $t\to\infty$,
and is not reliable when the cloning factor is larger than  $\mathcal{N}$ (shaded areas in Fig.~\ref{fig:SCGF_PA_ind} and~\ref{fig:SCGF_PA_const}),
as studied in~\cite{Hurtado2009}.
Our implementation  correctly reproduces the most relevant features of
the SCGF, i.e., the non-analyticity in $s_1$ and the constant branch for $s>s_1$, but
loses accuracy for large positive currents ($s < 0$) presumably due to the finite $\mathcal{N}$ effect.

In the PA process, the lowest eigenvalue does not appear to converge to a finite value in the limit $s \to \infty$.
From the condition~\eqref{eq:condition_s1_ind} for the eigenvalue crossover,
we suggest
\begin{equation}
    e(s)=
    \begin{cases}
        \frac{\a \b}{\b+\c}(1-e^{-s}) + \frac{\c \d}{\b+\c} (1-e^{s}),       &   s\le s_1 \\
            c  + \d (1-e^s) ,     &   s>s_1.
    \label{eq:SCGFind}
    \end{cases} 
\end{equation}
The right branch can be physically understood by separating the contributions of the
particles leaving the site rightwards, which contribute a term $c$ as in the TA case,
and the particles injected from the right boundary, which independently follow a Poisson
process with rate~$\d$ and contribute a term $\d(1-e^{s})$.
Since in this regime the particles pile up,
the corresponding SCGF branch does not depend on the left boundary.
Numerical simulations, shown in Fig.~\ref{fig:SCGF_PA_ind}(b), confirm our argument.
There is no analogue, in the memory-less ZRP, of the $c$-dependent dynamical phase for $s>s_1$,
which arises as a consequence of the temporal correlations.

\begin{figure}
 \centering
   \includegraphics[width=8cm]{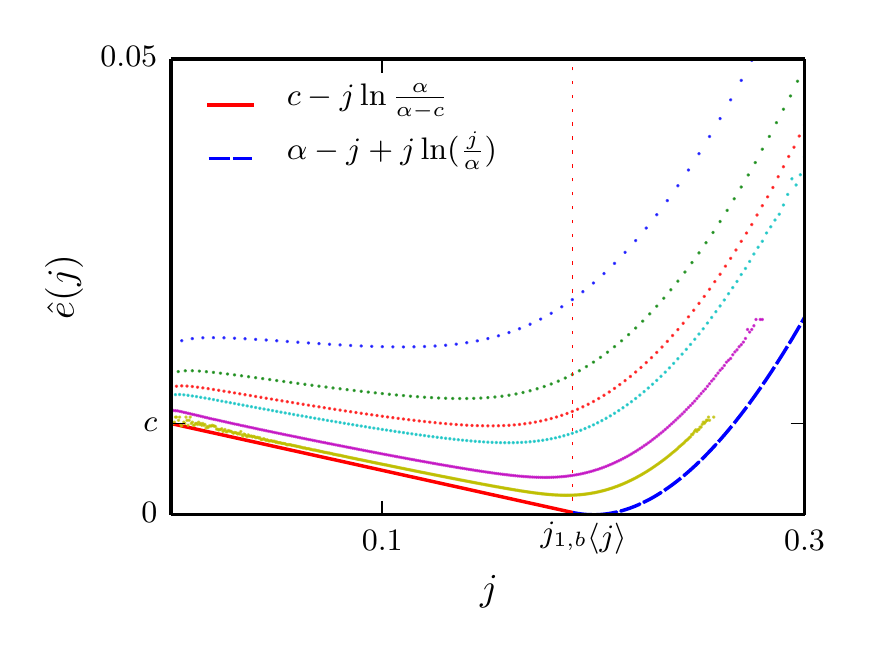}
   \caption{\label{fig:rate_function_ind}
    (Color online)
    Rate function $\hat{e}(j)$ for the on-off TA process with $(\a,\b,\g,\d,c)=(0.2,0.3,0,0,0.01)$ and $\mu_n=n$ (solid line).
    The points are simulation  data for the finite-time rate function $-\ln[\Prob(J/t=j)]/t$ computed at 
    $t=200,300,400,500,600,700$ (top to bottom).
    }
\end{figure}

For SCGFs with non differentiable points, as in Eq.~\eqref{eq:SCGFind},
the Legendre-Fenchel transform~\eqref{eq:legendre} of $e(s)$ gives in general the convex hull of the rate function $\hat{e}(j)$,
which can hide a non-convex shape.
However, for this system, we argue on physical grounds (see following) that Eq.~\eqref{eq:legendre}
gives indeed the true rate function, i.e.,
{\small
\begin{equation}
\begin{aligned}
&\mathlarger{\hat{e}(j)}\\
&=    \begin{cases}
        c + \d + j - j \ln(\frac{-j}{\d}), & \text{ } j \leq j_{1,a}\\
        -s_1 j + c + \d(1 -e^{s_1}),        & \text{ } j_{1,a} <j<  j_{1,b}\\
         \begin{gathered}
    \frac{\a \b}{\b + \g} + \frac{\g \d}{\b + \g}  - \sqrt{j^2 + 4 \frac{\a \b}{\b + \g} \frac{\g \d}{\b + \g}}   \\
    + j \ln \frac{j+\sqrt{j ^2 + 4 \frac{\a \b}{\b + \g}  \frac{\g \d}{\b + \g}}}{2 \frac{\a \b}{\b + \g}} ,
         \end{gathered}      & \text{ } j \geq j_{1,b}.
    \end{cases}
\end{aligned}
\end{equation}}
The two critical currents $j_{1,a} = -\d e^{s_1} $ and $j_{1,b}=\frac{\a \b}{\b + \g} e^{-s_1}-\frac{\g \d}{\b + \g}e^{s_1}$ are,
respectively, the right and left derivatives of $e(s)$ at $s=s_1$. In the TA process $j_{1,a}=0$.
The phase $j \leq j_{1,a}$ is obtained from the Legendre-Fenchel transform of $e(s)$ in the interval $s>s_1$,
while the phase $j \geq j_{1,b}$ is derived from $e(s)$, with $s<s_1$.
The transition value $s_1$ is mapped to the linear branch in $j_{1,a}<j<j_{1,b}$. 
This behaviour is equivalent to an ordinary equilibrium first-order phase transition,
where a linear branch of a thermodynamic potential corresponds to the coexistence of two phases.
In this non-equilibrium system, the mixed phase consists in a regime where,
for some finite fraction of time,
the current assumes value $j_{1,a}$, while for the rest of the time it has value $j_{1,b}$.  
As a result, the rate function in this region is linear with $j$, as predicted
by the Legendre-Fenchel transform.
This argument is supported by standard Monte Carlo simulations (ensemble size of $10^{10}$),
and it is particularly evident in the TA case (Fig.~\ref{fig:rate_function_ind}).

The different phases can be physically understood by observing the effect of the particle blockade.
In the case  with TA hopping rates, when the site is {\small OFF}, the particles accumulate and the outgoing current is necessarily zero.
The zero current is mapped to the flat section of the SCGF.
This is the dominant mechanism responsible for zero current.
At the end of an \OFF period, we have a configuration with many particles on the site.
When the lock is released, particles can leave the site with a rate proportional
to the occupation number. 
Consequently, the particles are quickly released after an \OFF period and the current jumps to a positive value.
In particular, the probability of having currents larger than $j_{1,b}$ is dominated by the phases in which the site is~{\small ON}.
In the presence of arrivals from the right boundary ($\d \neq 0$),
the blocked configuration becomes important for negative currents $j<j_{1,a}$,
and the rate function has an additional term corresponding to an independent Poisson process with  rate~$\d$.

As an aside, the dynamical phase transition seen at $s_1$ is not restricted to the
particular on-off ZRP explored here.
For example, an alternative on-off ZRP with unbounded departure rates
and on-off dynamics independent from the arrivals, displays the same fluctuating scenario.
Also, spatially extended spin systems such as the contact process~\cite{Lecomte2007}
and some kinetically constrained models~\cite{Garrahan2007},
can possess active and inactive phases coexisting at $s=0$. 

\noindent \textbf{Constant departure rates.}
In this case when $\m_n=1$, $n>0$, the operator $\tilde{H}$  has a continuous 
band which governs the fluctuations in certain regimes.
A way to obtain the SCGF is to evaluate the long-time limit of the
matrix element $\langle 1| e^{-\til{H} t}| P(0) \rangle$
by computing the full spectrum and the complete set of eigenvectors of $\til{H}$.
This task appears to be rather complicated for the $s$-modified Hamiltonian \eqref{eq:biased_hamiltonian},
requiring spectral theory and integral representation of block non-stochastic
operators~\cite{*[{The integral representation of Markov chains described by \textit{stochastic}
block tridiagonal generators is derived for example in~}] [{.}] Dette2007}.
As an approximation, we can use the reduced operator~\eqref{eq:operator}
and study the simpler expectation $\langle 1|e^{-\til{H}^{\star} t}  | P(0) \rangle$.
Recall that $\til{H}^{\star}$ has the same lowest eigenvalue $A_0$ as $\til{H}$,
at least in the regime  $s_2 \leq s \leq s_1$ where the ansatz~\eqref{eq:ansatz_general1}--\eqref{eq:ansatz_general2}
is valid.
Outside this regime it is expected to yield only approximate information about the current fluctuations.

The integral representation allows us to take into account the
dependence of the fluctuations on the initial condition.
We follow the same procedure as~\cite{Harris2006,Rakos2008},
with the difference that the departure rate here depends on $s$.
In fact, the solution found only has a weak dependence on the functional form of $w_{c,s}$
but, nevertheless, we report the explicit calculations for completeness.
As initial condition, we choose a geometric distribution with parameter $x$, i.e.,
$| P(0) \rangle = (1-x) \sum_{n=0}^\infty x^n | n \rangle $ where $|n\rangle$
denotes the configuration of the site with $n$ particles and
is an element of the natural basis for $\til{H}^{\star}$. The steady state is obtained
for $x=z w^{-1}_c$, where $w_c=\m(\a+\d+c)/(\a+\d+c+(\b+\g)\m)$ and $z=(\a+\d)/(\b+\g)$ are
the PA counterparts of the effective departure rate and fugacity found in Sec.~\ref{sec:stationary_distribution},
while the limit $x \to 0$ corresponds to the empty-site state.
The exact calculation of the full spectrum and of its eigenvectors, reported in Appendix~\ref{sec:spectrum},
gives the following representation:
\begin{multline}
 \langle 1|e^{-\til{H}^{\star} t}|P(0) \rangle  =  -\frac{1-x}{2 \pi i x \phi } \oint_{C_1} \frac{e^{- \varepsilon(\zeta) t }}{(\zeta-\frac{1}{x \phi})(\zeta - \frac{1}{\phi})}d\zeta    \\
 - \frac{1-x}{2 \pi i x  } \oint_{C_2} \frac{(y \zeta -1) e^{-\varepsilon(\zeta) t} }{(\zeta - \frac{1}{\zeta \phi})(\zeta - \phi)(\zeta-y)}d\zeta,
 \label{eq:integral}
\end{multline}
where $\varepsilon(\zeta)$ is obtained from the expression for the continuous band of the spectrum $\epsilon(k)$
after the substitution $\zeta = e^{ik}$ and
\begin{align}
\phi = &\sqrt{\frac{ (\beta e^{-s} + \gamma) w_{c,s} }{(\alpha + \delta e^{s})}} , \label{eq:phi} \\
y  = &\frac{1}{(\b+\g)w_{c,s}} \sqrt{(\alpha + \delta e^{s}) (\beta e^{-s} + \gamma) w_{c,s}} ,  \label{eq:y} \\
\epsilon(k) =& \a + \d + (\b + \g) w_{c,s} \notag \\
&- 2 \sqrt{(\a  + \d e^{s})(\b e^{-s} + \g)w_{c,s}} \cos(k).  \label{eq:epsilon}
\end{align}
The integration contours $C_1$ and $C_2$ are anti-clockwise circles centred around the origin with radius
$\phi^{-1}<\lvert  \zeta \rvert < (\phi x)^{-1}$ 
and infinitesimal size respectively.

The long-time limit of this integral is computed by means of the method of steepest descents
with saddle point at $\zeta = 1$.
When the saddle point contour engulfs one of the poles of the integrand we must also take
into account the residue~\cite{Touchette2010}.

\begin{figure}
 \centering
    \includegraphics[width=8cm]{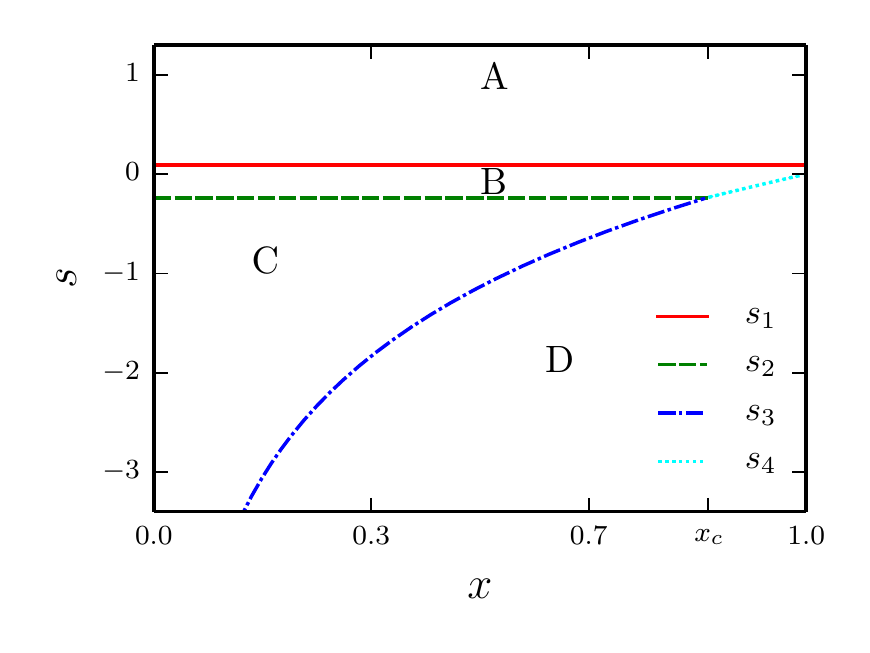}
 \caption{\label{fig:phase_diagram} (Color online) Phase diagram, based on Eq.~\eqref{eq:integral},
                for the current fluctuations in the PA process with $\m_n=1$
                and $(\a,\b,\g,\d,c)=(0.1,0.2,0.1,0.1.0.5)$.
                The lines $s_1$ and $s_4$ correspond to first-order
                dynamical phase transitions while $s_2$ and $s_3$ mark second-order transitions. }
\end{figure}

For  fixed parameters $s$ and $x$, the leading term in the long-time limit of $\langle 1|e^{-\til{H}^{\star} t}|P(0) \rangle $ is given by the slowest
decaying exponential and the SCGF is determined by one of the rates $\varepsilon(\phi),\varepsilon(y),\varepsilon(1),\varepsilon((x \phi)^{-1})$.
Tuning $s$ or $x$, the positions of the poles with respect to the saddle point contour are altered
and the leading term in the integral expansion changes.
This produces the phase diagram of Fig.~\ref{fig:phase_diagram} for the SCGF.
 The critical line corresponding to the solution of $\phi=y$ is $s=s_1$.
 The line $s=s_2$ corresponds to $y=1$. These two phase transitions were also found in Sec.~\ref{sec:range_of_validity}
 as critical points in the full state space.
 The curves $s=s_3$ and $s=s_4$ are solutions of $1=(\phi x)^{-1}$ 
 and $y=(\phi x)^{-1}$ respectively.
 The tri-critical point $s_3 = s_4$ is at $x_c$.
It is worth noting that higher positive current fluctuations retain a dependence on the initial condition $x$
and that, unlike the memoryless ZRP, the critical point $s_1$ can fall in the positive current range. 
The explicit expressions in terms of $s$ for the TA case are reported in Appendix~\ref{sec:phase_TA}.

We distinguish four phases:
\begin{description}[leftmargin=0cm, font=\textit]
\item[Phase A]  $s>s_1$. In this case the leading term arises from the pole at $\zeta=\phi$.
The product $\langle 1 | \til{P}_{A_0}\rangle$ diverges and the SCGF is different from the lowest eigenvalue $A_0$,
being given instead by
 \begin{equation}
    e(s) = \d  (1-e^s) +  \b w_{c,s} (1-e^{-s}).
    \label{eq:phaseA}
 \end{equation}
This phase corresponds to very small positive currents (in particular when $\delta=0$) or large backward currents.
Large negative currents are mainly governed by the rate $\d$ of particle arrival from the right,
which contributes to the SCGF with the first term of~\eqref{eq:phaseA}.
The second term corresponds to particles that jump rightwards from the site with an effective rate $\b w_{c,s}$.
The current fluctuations in this phase are optimally realised by a site with arbitrarily large occupation number (instantaneous condensation)
that acts as a reservoir, so that the outgoing current has no dependence on the left boundary hops~\cite{Rakos2008}.
We argue that the presence of a left and a right term in Eq.~\eqref{eq:phaseA} is generic for this phase,
although there is no \textit{a priori} reason for the effective rate $w_{c,s}$
to have the same form as in the small fluctuation regime.
In the PA case, for large values of $s$, the SCGF is dominated by the first term and is not
sensitive to the functional form of $w_{c,s}$.

\item[Phase B] $((s_2<s<s_1) \wedge (x<x_c)) \vee ((s_4<s<s_1) \wedge (x>x_c)))$.
 This phase arises when the pole at $\zeta=y$, corresponding to the lowest
 eigenvalue $A_0$ \eqref{eq:lowest_ev}, becomes dominant, hence
 \begin{equation}
    e(s) = \frac{\a \b}{\b+\c}(1-e^{-s}) + \frac{\c \d}{\b+\c} (1-e^{s}) .
    \label{eq:phaseB}
 \end{equation}
The probability of fluctuations in this regime is asymptotically identical to the standard ZRP.
In this range the site has finite occupation and the probability that a particle
leaves is conditioned to an arrival event, just as in~\cite{Juhasz2005,Harris2005,Rakos2008}.

 \item[Phase C] $(x< x_c)\wedge (s_3<s<s_2)$. This phase arises from the saddle-point at $\zeta=1$.
It corresponds to a large forward current sustained by a large inward current from the left boundary.
The asymptotic form \eqref{eq:naive} still holds,
but with an  oscillating (non-decaying in $n$) ground state.
This also represents an instantaneous condensate, but with particle number growing as the square root of time~\cite{Rakos2008}.
Here the spectrum of $\tilde{H}^{\star}$ is continuous and
the SCGF is given by the minimum of the band~\eqref{eq:epsilon}:
\begin{equation}
e(s) = \a + \d + (\b + \g)w_{c,s} - 2 \sqrt{(\a  + \d e^{s})(\b e^{-s} + \g)w_{c,s}}.
\label{eq:phaseC}
\end{equation}

\item[Phase D]   $[(s<s_3) \wedge (x<x_c)] \vee [(s<s_4) \wedge (x>x_c)]$.
This phase arises when the residue at $\zeta=(\phi x)^{-1}$
dominates the long-time behaviour:
\begin{multline}
e(s) = \alpha +\delta +  (\beta +\gamma ) w_{c,s} -  (\beta e^{-s} + \gamma)  w_{c,s}  x - (\alpha + \delta e^s)/x.
\label{eq:phaseD}
\end{multline}
It corresponds to a large forward current of particles that
is most likely to be realized from an initial configuration with very high occupation number
and also has an analogue in the standard ZRP~\cite{Rakos2008}.
\end{description}

These results are compared to the cloning simulations in Fig.~\ref{fig:SCGF_PA_const}  for $x \to 0$.
Similarly to the independent-particle case, the cloning data for the left branch,
corresponding to large positive currents, is potentially affected by finite-$\mathcal{N}$ effects~\cite{Hurtado2009}.
It turns out that for the chosen parameters our approximation~\eqref{eq:phaseA}--\eqref{eq:phaseC}, plotted as a solid line,
is very close to the naive approach (not shown) in which the same representation~\eqref{eq:integral} is used,
but the effective departure rate has the $s$-independent form $w_c$~(see Sec.~\ref{sec:stationary_distribution}) for all the regimes.
The analytical SCGF does not match the simulation points in either of the phases A and C.
We attribute this to the failure of the assumption~\eqref{eq:ansatz_general1}--\eqref{eq:ansatz_general2} for the ground state in phases A and C.
In other words, large fluctuations cannot be exactly described by an effective departure rate $w_{c,s}$
with a simple functional dependence on $s$.

\begin{figure}
 \centering
 \includegraphics[width=8cm]{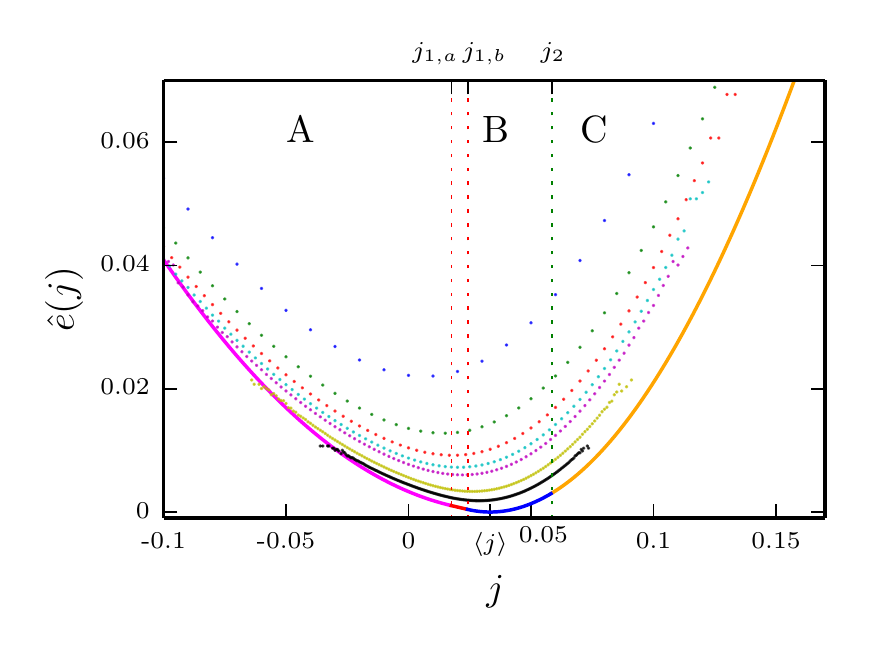}
 \caption{\label{fig:rate_function_const} (Color online)
 Rate function for the on-off ZRP with $(\a,\b,\g,\d,c)=(0.1,0,2,0.1,0.1,0.5)$ and $\mu_n=1$. 
 Points are data for $-\ln[\Prob(J/t = j)]/t$ from standard Monte Carlo simulation at times $t=100,200,300,400,500,1000,2000$.
 (top to bottom).  The solid line is the analytical approximation for $t \to \infty$.
}
\end{figure}

In Fig.~\ref{fig:rate_function_const}, the rate function $\hat{e}(j)$,
computed by means of a Legendre-Fenchel transform on the SCGF~\eqref{eq:phaseA}--\eqref{eq:phaseC},
is compared to the finite-time rate function obtained from standard Monte Carlo simulations with an ensemble size of $10^{10}$.
Although approximate, $\hat{e}(j)$ appears to capture well the shape of the long time limit for the simulation data points.

\subsection{Numerical results for large system}
\begin{figure}
    \centering
    \includegraphics[width=8cm]{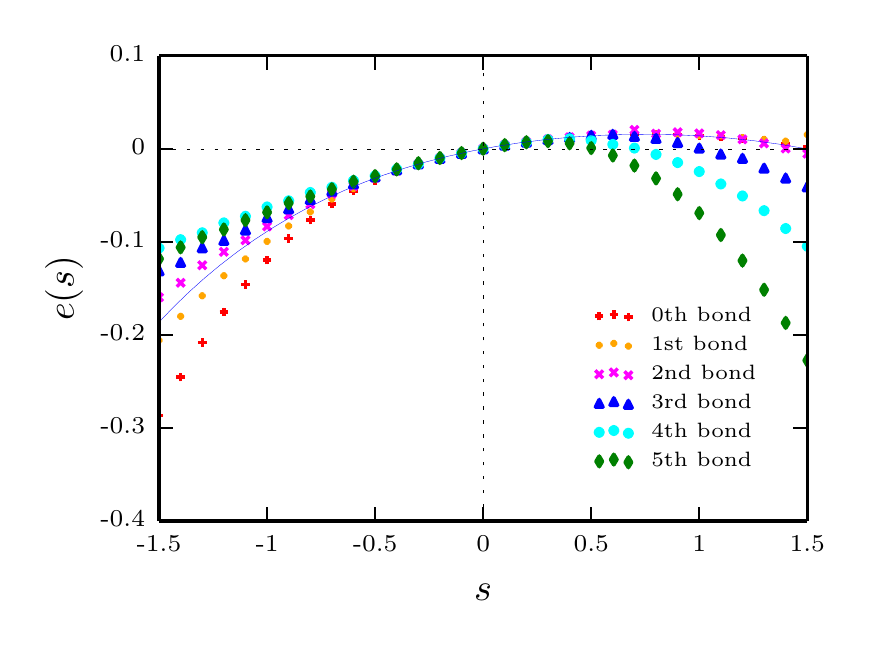}
    \caption{\label{fig:SCGF_PA_const_chain} (Color online) Simulation results for the SCGF
    in a five-site on-off ZRP with $\m_n=1$ and
    $(\a,\b,\g,\d,p,q,c)=(0.1,0.2,0.1,0.1,0.55,0.45,0.5)$.
    The solid line is the $c$-independent expression for the
    lowest eigenvalue of the $s$-modified Hamiltonian for the five-site Markovian-ZRP~\cite{Rakos2008}.
    }
\end{figure}

The lack of a stationary product form solution for the on-off ZRP on an extended lattice
makes the analytical study of fluctuations, across the generic bond, impractical.
It would be possible to use the mean-field stationary solution to derive an approximate
SCGF using the same procedure as in the single-site model.
However, we do not expect the result to be accurate, especially for small values of $c$ and for current fluctuations far from the mean.
To explore the larger system we make use of the cloning method, see Fig.~\ref{fig:SCGF_PA_const_chain}.
While the statistics of rare currents is bond-dependent, it is possible to appreciate that for each
bond the SCGF matches that of a Markovian ZRP in the neighbourhood of $s=0$, a feature shared with the one-site system.

The central regime satisfies a Gallavotti-Cohen fluctuation symmetry~\cite{Lebowitz1999}
$e(s)=e(E-s)$, with $E=\ln[ (p/q)^{L-1} \a \b/ (\g \d)]$.
Such a relation seems to be ensured by the fact that the relative probabilities
of particle jumps towards the left or the right are independent of the time that the particle spends on a site.
This property is related to the \textit{direction--time independence} of Ref.~\cite{Andrieux2008}.
However, the fluctuation symmetry is not guaranteed to hold on an arbitrary domain
in systems with infinite state space~\cite{Harris2006}.
In fact, as expected, we see here a $c$-dependent breakdown for large fluctuations.

\section{Discussion}
\label{sec:conclusions}
We have studied an open-boundary zero-range process
that incorporates memory by means of an additional ``phase'' variable.
The particles are blocked on a lattice site (``phase \off'') when a new particle arrives
and consequently congestion is facilitated.
After an exponentially distributed waiting time with parameter $c$, the block is removed (``phase \on'').
At first sight, the effects of time correlations are hidden.
The stationary state solution of the one-site system can be written as in
the Markovian case, with an effective on-site interaction $w_{c,n}$.
This means that, if the direct interactions are unknown,
it is not possible to distinguish a single site with on-off dynamics from
a standard memoryless ZRP by looking only at the occupation distribution.

However, the presence of \ON and \OFF phases alters the statistics of the outwards particle hops.
This becomes important in the spatially extended system
where each site receives particles, from its neighbours,
according to a non-Markovian process.
As a consequence, a product form solution is in general not expected
and we have relied on a mean-field approach for the analytical treatment.
This approximation consists of replacing the true particle
arrival on each site with a memoryless process, while keeping
exact information about the on-site particle departure as well as the lattice topology.
This procedure can be applied in principle to decouple non-Markovian ZRPs on an arbitrary lattice,
provided that it is possible to solve the consistency equation for the fugacities.
We found that, in the chain topology studied here, the mean-field approach is very accurate for large values of $c$ and gives an
analytical estimate $c_{\mathrm{mf}}$ for the congestion threshold.

The memory effects at the fluctuating level appear more interesting even in the single-site case.
Fluctuations close to the mean current are obtained by analytic continuation of the stationary
state and are indistinguishable from the fluctuations in a memoryless ZRP.
However, under certain conditions 
large current fluctuations are optimally realized by the instantaneous piling
up of particles on the site and the statistics of such fluctuations change abruptly.
In the absence of direct inter-particle interaction we have found a memory-induced
dynamical first-order phase transition, i.e., the scaled cumulant generating function (SCGF) $e(s)$ is
non-analytic at a particular value $s_1$.
In the totally asymmetric case, this occurs only if the parameter $c$ is smaller than the arrival rate $\a$.
The system with constant departure rates, i.e., attractive inter-particle interaction,
undergoes second-order as well as first-order dynamical phase transitions.
The state of the system during a small fluctuation event has the same form as the stationary state,
but with a more general modified effective interaction factor.
Indeed, the exact phase boundaries and the large deviation function of this regime
are encoded in the reduced operator $\til{H}^\star$ [Eq.~\eqref{eq:operator}],
which has the same structure as 
the $s$-modified Hamiltonian of the standard ZRP,
but with an $s$-dependent effective interaction factor.
We have used the same operator $\til{H}^\star$ to find an approximate solution for the fluctuations outside this phase.
Numerical tests confirm the presence of the predicted $c$-dependent dynamical phase transitions.

The separation between a small-fluctuation regime, with a memory independent SCGF, and
high-fluctuation regimes, where memory plays a more obvious role, is
a feature also found numerically in the spatially-extended system.
It would be of interest to explore the role of topology in more detail
as well as to look for similar memory effects in other driven interacting-particle systems.
Furthermore, we point out the importance of solving the eigenproblem~\eqref{eq:eigenproblem} for the full
$s$-modified Hamiltonian $\tilde{H}$ [Eq.~\eqref{eq:biased_hamiltonian}]
which provides exact information about the strongly fluctuating regimes.
This would be of interest in queueing theory; in fact quasi-birth-death processes,
which contain as a special case the single-site on-off model studied here, are widely used for
performance modelling of non-Markovian systems~\cite{Neuts1981,Stewart2009probability}.

To conclude, for the model explored in this paper, time correlations can be absorbed in an effective
memoryless description for the steady state, but can emerge at the fluctuating level and alter the probability of observing rare phenomena.
Such an observation leaves interesting open questions about the predictive power of effective
theories for real-world systems, where rare events can be of crucial importance.

\begin{acknowledgments}
 We thank Pablo Hurtado for useful discussion about the cloning algorithm.
 In addition, RJH is grateful for the hospitality of the National Institute for Theoretical Physics
 (NITheP) Stellenbosch during the final stages of this work.  
 The research utilised Queen Mary's MidPlus computational facilities,
 supported by QMUL Research-IT and funded by EPSRC Grant No. EP/K000128/1.
\end{acknowledgments}

\appendix

\section{Derivation of the stationary state}
\label{sec:stationary_on_off}
Summing Eqs.~\eqref{eq:1} and~\eqref{eq:3}, and imposing the stationarity condition, it follows that
\begin{multline}
	\b \mu_{n+1}  P^*(n+1,\on)  - \a P^*(n ) \\
	=\b \mu_{n}  P^*(n,\on)  - \a P^*(n-1),
	\label{eq:hirsch_onoff_A}
\end{multline}
while the stationarity conditions on Eqs.~\eqref{eq:2} and~\eqref{eq:4} imply the boundary conditions
\begin{gather}
    \b \mu_1  P^*(1,\on) - \a P^*(0,\on) = 0,     \label{eq:hirsch_onoff_B}\\
	P(0,\off)=0 ,
\end{gather}
which, together with~\eqref{eq:hirsch_onoff_A}, allow us to write the recursive relation
\begin{gather}
	\b \mu_{n+1} P^*(n+1,\on) = \a P^*(n , \on) + \a P^*(n, \off).
	\label{eq:hirsch_recursive}
\end{gather}
Using the stationarity condition on Eq.~\eqref{eq:3}
\begin{gather}
	 (\a + c) P^*(n+1, \off) = \a P^*(n, \on) + \a P^*(n, \off),
\label{eq:hirsch_onoff_C} 
\end{gather}
we eliminate $P(n)$ from Eqs.~\eqref{eq:hirsch_recursive} and~\eqref{eq:hirsch_onoff_C} and get
\begin{equation}
	(\a + c) P^*(n+1, \off) =   \b \mu_{n+1} P^*(n+1,\on), \label{eq:ciaos} 
\end{equation}
hence,
\begin{align}
	P^*(n,\off) &= \frac{\b \mu_n }{\a + c +  \b \mu_n}P^*(n) , \\
	P^*(n,\on) &= \frac{(\a +c )}{\a + c +  \b \mu_n} P^*(n).  
\end{align}
The ratios $(\b \m_n )/(\a + c +  \m_n \b)$ and $(\a +c )/(\a + c + \b \m_n )$
are the conditional probabilities $P^*(\off|n)$ and $P^*(\on|n)$, respectively.
Substituting in \eqref{eq:hirsch_recursive} or \eqref{eq:hirsch_onoff_C}
we get the recursive relation:
\begin{gather*}
	\frac{ \m_n (\a + c)}{\a + c + \b \m_n }  P^*(n+1) =  \frac{\a}{\b } P^*(n).
\end{gather*}
Finally, iterating and using the definitions of $z$ and $Z_c$ we find the probability mass~\eqref{eq:stationary_on_off_1}.

\section{Left eigenvectors of the $s$-modified Hamiltonian}
\label{sec:left_eigen_ind}
The derivation of $\langle \tilde{P}_{A_0} |$ when $\mu_n=\mu$, $n>0$, is as follows.
Assuming that the left-eigenvector components satisfy
\begin{align}
    P_{\text{left}}(n,\on) &= p_{\text{\tiny ON},s}^{\text{\tiny left}} P_{\text{left}}(n), \label{eq:ansatz_left1}\\
    P_{\text{left}}(n,\off) &= (1-p_{\text{\tiny ON},s}^{\text{\tiny left}}) P_{\text{left}}(n), \label{eq:ansatz_left2}\\
    P_{\text{left}}(n+1) &= {\rho^{\text{\tiny left}}_s} P_{\text{left}}(n),    \label{eq:ansatz_left3}
\end{align}
we get the explicit equations
\begin{align}
\begin{multlined}
    -(\alpha -A+c+\delta ) (1-p_{\mbox{\tiny ON},s,0}^{\text{\tiny left}})  + c p_{\mbox{\tiny ON},s,0}^{\text{\tiny left}} \\
    + (\alpha +\delta  e^s) {\rho^{\text{\tiny left}}_s}  (1-p_{\mbox{\tiny ON},s}^{\text{\tiny left}}) = 0,
\end{multlined}& \label{eq:B1}\\
\begin{multlined}
     (\alpha +\delta  e^s) \rho^{\text{\tiny left}}_s  (1-p_{\mbox{\tiny ON},s}^{\text{\tiny left}}) -  (\alpha -A+\delta ) p_{\mbox{\tiny ON},s,0}^{\text{\tiny left}} = 0,  \label{eq:B2} 
\end{multlined}&\\
\begin{multlined}
    - (\alpha -A+c+\delta )  (1-p_{\mbox{\tiny ON},s}^{\text{\tiny left}})  + c \left(  1-p_{\mbox{\tiny ON},s}^{\text{\tiny left}}  \right)\\
    +  \left(\alpha +\delta  e^s\right)  \rho^{\text{\tiny left}}_s (1-p_{\mbox{\tiny ON},s}^{\text{\tiny left}}) = 0, 
\end{multlined}&  \label{eq:B3}
\end{align}
\begin{align}
 \begin{split} 
    (\beta e^{-s}+\gamma) \mu p_{\mbox{\tiny ON},s,0}^{\text{\tiny left}} &- (\alpha -A+(\beta +\gamma)\mu+\delta)\rho^{\text{\tiny left}}_s p_{\mbox{\tiny ON},s}^{\text{\tiny left}} \\
    &+  \left(\alpha +\delta  e^s\right) {\rho^{\text{\tiny left}}_s}^2 (1-p_{\mbox{\tiny ON},s}^{\text{\tiny left}}) = 0 ,
\end{split}&  \label{eq:B4}\\
\begin{split}
    -{\rho^{\text{\tiny left}}_s} p_{\mbox{\tiny ON},s}^{\text{\tiny left}} [\alpha -A+&\mu  (\beta +\gamma )+\delta ]
    +\mu  p_{\mbox{\tiny ON},s}^{\text{\tiny left}} \left(\gamma +\beta  e^{-s}\right)\\
    &+{\rho^{\text{\tiny left}}_s}^2 (1-p_{\mbox{\tiny ON},s}^{\text{\tiny left}}) \left(\alpha +\delta  e^s\right) = 0 ,
\end{split}& \label{eq:B6}
\end{align}
where the factor $p_{\mbox{\tiny ON},s,0}^{\text{\tiny left}}$ is assumed to be different from $p_{\mbox{\tiny ON},s}^{\text{\tiny left}}$ by analogy with the right eigenproblem.
The Eqs.~\eqref{eq:B1} and~\eqref{eq:B2} give
$(\alpha -A+c+\delta) (1- p_{\mbox{\tiny ON},s,0}^{\text{\tiny left}})+(\alpha -A+\delta)p_{\mbox{\tiny ON},s,0}^{\text{\tiny left}} + c p_{\mbox{\tiny ON},s,0}^{\text{\tiny left}} = 0$,
which is verified for $p_{\mbox{\tiny ON},s,0}^{\text{\tiny left}}=\frac{1}{2}$.
The Eqs.~\eqref{eq:B4} and~\eqref{eq:B6} imply $p_{\mbox{\tiny ON},s}^{\text{\tiny left}}=p_{\mbox{\tiny ON},s,0}^{\text{\tiny left}}$.
After the substitution, the remaining equations are solved for $A=A_0$ and ${\rho^{\text{\tiny left}}_s} = (\beta  e^{-s}+\gamma )/(\beta +\gamma )$.
With those constants, it is easy to verify that the ansatz~\eqref{eq:ansatz_left1}--\eqref{eq:ansatz_left3} is consistent even in the general departure rate case.
In fact, after substitution, all the terms containing $\mu_n$ cancel out.
In the reduced state space we get a consistent result since the row vector $\langle P^{\star}_{A_0}|$ with components given by~\eqref{eq:ansatz_left3}
satisfies $\langle P^{\star}_{A_0}| \til{H}^{\star} = A_0 \langle P^{\star}_{A_0}|$.

\section{Spectrum and integral representation}
\label{sec:spectrum}
In this appendix, we report the calculations which lead to the integral representation~\eqref{eq:integral}.
Let us impose an initial condition of Boltzmann type for the system, so that
\begin{equation}
 \langle 1| e^{- \tilde{H}^\star t}| P(0)\rangle = ( 1- x) \sum_{n,m=0}^{\infty} x^n  \langle m | e^{- \tilde{H}^\star t} |n\rangle,
 \label{eq:gen_IC}
\end{equation}
where $\langle m |$ ($|n\rangle$) is a row (column) vector with a ``$1$'' in the $m$-th ($n$-th) position and ``$0$'' elsewhere.
To evaluate the right-hand side of~\eqref{eq:gen_IC}, we first seek for normal modes of the dynamics generated by the operator $\tilde{H}^\star$~\eqref{eq:operator}.
We transform $\tilde{H}^\star$ into the symmetric form $\Phi \tilde{H}^\star \Phi^{-1}$, where
$\Phi$ is the diagonal operator with entries $\delta_{ij} \phi^{i}$, $\delta_{ij}$ is the Kronecker delta, $i,j =0,1,2,\ldots$,
and $\phi$ is the combination of parameters~\eqref{eq:phi} in the main text.
The associated eigenproblem is solved after a Fourier transformation.
Its eigenvalue $\epsilon(k)$~[Eq.~\eqref{eq:epsilon}],
has eigenvector $| \psi'(k) \rangle$ 
with components $\sqrt{2/\pi} \sin(n k + \varphi)$.
Substituting this in the first row equation for the eigenproblem, we get the following expression for~$\varphi$:
\begin{equation}
    e^{i2 \varphi } = \frac{1-e^{ik} y}{1-e^{-ik}  y},
    \label{eq:e2phi}
\end{equation}
where $y$ is given by the $s$-dependent expression~\eqref{eq:y}.
For $y<1$ a discrete eigenvalue appears with eigenvector $| \psi'(0) \rangle  = \sqrt{ 1- {y}^2  } \sum_{n=0}^{\infty} {y}^n |n \rangle $ and eigenvalue $A_0$~[Eq. \eqref{eq:lowest_ev}] while, for $y>1$, the infimum of the spectrum is given by $\epsilon(0)$.

The vectors $| \psi(k) \rangle = \Phi^{-1} | \psi'(k) \rangle $ , $k \in (0,\pi]$, and $| \psi(0) \rangle = \Phi^{-1} | \psi'(0) \rangle$ form
a complete set, i.e.\ $ \int_{0}^{\pi}| \psi(k) \rangle \langle \psi(k) | dk + | \psi(0) \rangle \langle \psi(0) | = \mathds{1} $.
Inserting this representation of the identity in Eq.~\eqref{eq:gen_IC},
the right-hand side becomes
\begin{multline}
  ( 1- x) \sum_{n,m=0}^{\infty} x^n \int_{0}^{\pi}  \langle m| \psi(k) \rangle \langle \psi(k) |n\rangle    e^{- \epsilon(k) t} dk  \\
+  \Theta(1-y) (1-x) \sum_{m,n=0}^{\infty} x^n \phi^{n-m} (1-{y}^2) y^{n+m} e^{-A_0 t},
\label{eq:integralE1}
\end{multline}
where $\Theta$ denotes the Heaviside step function.
Using the fact the  eigenvectors are odd in $k$, the integral in Eq.~\eqref{eq:integralE1} can be rewritten as
\begin{equation}
\int_{0}^{2\pi} \left( e^{ik(n-m)} - e^{ik(n+m)} e^{i2 \phi} \right)  e^{- \epsilon(k) t} dk  ,
\label{eq:integralE2}
\end{equation}
and, using Eq.~\eqref{eq:e2phi}, it becomes 
\begin{equation}
\oint_{|\zeta|=1} \left( \zeta^{n-m-1} -\zeta^{n+m}\frac{1-\zeta y}{\zeta - y} \right)  e^{- \varepsilon(\zeta) t} d\zeta,
\label{eq:integralE3}
\end{equation}
where $\zeta=e^{ik}$ and $\varepsilon(\zeta) = \epsilon[k(\zeta)]$. Deforming the integration contour to $C_1$ for the first term in the
integrand and to $C_2$ for the second term, we obtain the representation~\eqref{eq:integral}. The last term in Eq.~\eqref{eq:integralE1}
cancels out with a pole contribution at $\zeta=y$ for $y<1$.

\section{Phase diagram for the current fluctuations in the TA process with bounded departure rate}
\label{sec:phase_TA}
In this appendix, we report the analytical forms of the $c$-dependent transition lines between
the dynamical phases of the TA case with $\m_n = n$.
The resulting phase diagram is similar to the PA case (Fig.~\ref{fig:phase_diagram}), but
with the transition line identified  by $s_1$ mapped to a positive value of the current. 
\begin{description}[leftmargin=0cm]
\item[$\boldsymbol{s=s_1}$]
The knowledge of $| \tilde{P}_{A_0} \rangle$ is sufficient to verify when the pre-factor $\langle s |  \tilde{P}_{A_0} \rangle$ is finite,
i.e.,
\begin{gather}
  \frac{\a (c +e^{-s} \a + \b \mu)}{  (c  +e^{-s} \a) \b \mu }<1,\\
  s < s_1 =  \ln \left(\frac{\a(\b \mu-\a) }{c \a -c \b \mu +\a \b \mu} \right)\label{eq:cond3}.
\end{gather}
Notice that this condition makes sense when the denominator in the argument of the
logarithm in~\eqref{eq:cond3} is positive,
i.e., $c<\a \b \mu/(\b \m - \a)$, while the stationarity condition $\a < \b w_c$
ensures that the numerator is positive.
The phase boundary can also be obtained from solving $\phi=y$.

\item[$\boldsymbol{s=s_2}$]
This critical point marks the left boundary of the region where
the condition ${\rho^{\text{\tiny left}}_s} \rho_{s} < 1$ holds, i.e.,
\begin{gather}
 \frac{\a (c  +e^{-s} \a+ \b \mu)}{ (c  +e^{-s} \a ) \b \mu} e^{-s} < 1, \\
 s > s_2  =  -\ln \left(   \frac{\sqrt{c^2 + 4 c \m} - c}{2 \a} \right).
\end{gather}
It corresponds to a solution of $y=1$.

\item[$\boldsymbol{s=s_3}$]
This line corresponds to $(\phi x)^{-1}=1$.
The critical point $s_3$ satisfies 
\begin{equation}
e^{-s_3} = \frac{\sqrt{\left(c \mu  x^2 + \a^2\right)^2+4 \a ^2 \mu ^2 x^2}-c \mu  x^2+\a ^2}{2 \a  \mu  x^2}.
\end{equation}

\item[$\boldsymbol{s=s_4}$]
This phase boundary is $c$-independent, specifically
\begin{equation} 
    e^{-s_4} =   x^{-1}.
\end{equation}
It corresponds to the condition $y=(\phi x)^{-1}$.
\end{description}

\bibliography{FluctuationOnOffZRP}

\end{document}